\documentclass[aps,prd,a4paper,preprintnumbers,nofootinbib,superscriptaddress, notitlepage]{revtex4}
\pdfoutput=1
\usepackage{latexsym}
\usepackage{amsthm}
\usepackage{amsmath}
\usepackage[dvips]{graphicx}
\usepackage{wrapfig}
\usepackage{amssymb}
\usepackage{color}
\usepackage{cancel}
\usepackage{multirow}
\usepackage{pst-3d}                  % PSTricks 3D framed boxes
\usepackage{pst-grad}                % PSTricks gradient mode
\usepackage{pst-node}                % PSTricks nodes
\usepackage{pst-slpe}                % Improved PSTricks gradients
\usepackage{pst-plot}                %
\usepackage{pst-coil}                %
\usepackage{pst-math}                %
\usepackage{pstricks-add}            %
\usepackage{rotating}
\usepackage{geometry}
\usepackage{hhline}
\usepackage{amsmath}
\usepackage{float}
\usepackage{ulem}
\usepackage{fancybox}

\usepackage[colorlinks=true,citecolor=darkred,urlcolor=darkred, pdfborder={0 0 0}]{hyperref}
\definecolor{darkred}{rgb}{0.6,0,0}
\usepackage{braket}

\allowdisplaybreaks
\allowdisplaybreaks[1]
\allowdisplaybreaks[2]

\definecolor{lightred}{rgb}{1,0.4,0.4}
\definecolor{lightgreen}{rgb}{0.4,1,0.4}

\allowdisplaybreaks
\allowdisplaybreaks[1]
\allowdisplaybreaks[2]

\begin{document}

\title{Confronting Tri-direct CP-symmetry models to neutrino oscillation experiments}
\author{Gui-Jun Ding\footnote{Email: dinggj@ustc.edu.cn}}
\affiliation{Interdisciplinary Center for Theoretical Study and Department of Modern Physics,\\University of Science and Technology of China, Hefei, Anhui 230026, China}
\author{Yu-Feng Li\footnote{Email: liyufeng@ihep.ac.cn}}
\affiliation{Institute of High Energy Physics, and School of Physical Sciences, University of Chinese Academy of Sciences, Beijing 100049, China}
\author{Jian Tang\footnote{Corresponding author, email: tangjian5@mail.sysu.edu.cn}}
\author{Tse-Chun Wang\footnote{Email: wangzejun@mail.sysu.edu.cn}}
\affiliation{School of Physics, Sun Yat-Sen University, Guangzhou 510275, China}
\date{\today}

\begin{abstract}

Tri-direct CP symmetry is an economical neutrino model building paradigm, and it allows red for the description of neutrino masses, mixing angles and CP violation phases in terms of four free parameters. Viability of a class of tri-direct CP models is examined with a comprehensive simulation of current and future neutrino oscillation experiments. The full parameter space of four independent parameters is carefully scanned, and the problem of parameter degeneracy appears for the constraints from one group of neutrino oscillation experiments. Two benchmark models which are promising from a model building point of view are also examined. Complementary roles from accelerator neutrino experiments (e.g., T2HK and DUNE) and reactor neutrino experiments (e.g., JUNO) are crucial to break the degeneracy and nail down the fundamental neutrino mixing parameters of the underlying theory.

\end{abstract}

% \preprint{USTC-ICTS-19-10}

\maketitle

\section{Introduction}
Neutrinos in the standard model (SM) of particle physics are strictly massless. Neutrino oscillation requires mass-squared differences and non-zero neutrino masses, which is striking new physics beyond the SM and calls for new degrees of freedom. In the framework of the three-generation neutrino oscillation paradigm, we have two mass-squared differences ($\Delta m_{21}^2$, $\Delta m_{31}^2$), three mixing angles ($\theta_{12}$, $\theta_{13}$ and $\theta_{23}$) and Dirac CP phase $\delta_\mathrm{CP}$ \cite{Esteban:2018azc}. The precision of measuring $\theta_{13}$ is dominated by reactor neutrino experiments~\cite{Adey:2018zwh,DoubleChooz:2019qbj,Abe:2018wpn}, $\theta_{12}$ and $\Delta m_{21}^2$ are dominated by solar and reactor neutrino experiment KamLAND~\cite{Cleveland:1998nv,Abe:2010hy,Aharmim:2011vm,Ghiano:2019cjy}, as well as $\theta_{23}$ and $|\Delta m_{31}^2|$ are dominated by atmospheric neutrino experiments~\cite{Abe:2018wpn,NOvA:2018gge,Adamson:2013ue}. A global analysis of different experiments provides the precise values of mixing parameters at the percentage level~\cite{Esteban:2018azc}. However, the mass ordering $\Delta m_{31}^2>0$ or $\Delta m_{31}^2<0$ and the value of $\delta_\mathrm{CP}$ remains unclear, although hints exist from experiments which are currently running.

Many models have been proposed to accommodate massive neutrinos without violating overwhelming constraints from previous experimental results. The origin of neutrino masses, flavor mixing and CP violation is a longstanding open question in particle physics. It turns out that a broken flavour symmetry based on a discrete group is particularly suitable to explain the structure of the leptonic mixing matrix, see Refs.~\cite{Altarelli:2010gt,Ishimori:2010au,King:2013eh,King:2014nza,King:2015aea} for review. If the discrete flavor symmetry is extended to {also involve} CP as a symmetry, the CP violation phases in the quark sector (observed) and lepton sector can be predicted~\cite{Li:2017abz,Lu:2018oxc,Lu:2019gqp}. Recently a new discrete flavor symmetry model building approach called tri-direct CP was proposed~\cite{Ding:2018fyz,Ding:2018tuj}, and it is dictated by residual symmetries such that it is quite predictive. The light neutrino mass matrix only depends on four real free parameters to describe the entire neutrino sector (three neutrino masses as well as the lepton mixing matrix). Moreover the CP violations in neutrino oscillations and leptogenesis generally arise from the same phase in the tri-direct CP model, consequently they are closely related to each other.

Precision measurements of neutrino oscillation parameters will guide us to the new physics domain. While in the quark sector, the precision is at the sub-{percentage} level~\cite{Tanabashi:2018oca}, in the neutrino sector, the parameter uncertainties remain at the {percentage} level~\cite{Esteban:2018azc}. New physics might be hidden in the uncertainties of measured neutrino mixing parameters, as neutrino oscillations bridge neutrino mixings and other factors which affect the propagation of coherent states. The new models to accommodate massive neutrinos intend to bring new fundamental symmetries, new particles and their new interactions beyond {the} standard model. It is promising to conduct precision measurements {in} accelerator neutrino experiments to search for new physics, including non-standard interactions, and neutrino decays (\textit{e.g.}~Ref.~\cite{Wang:2018dwk,Esteban:2018ppq,Tang:2017khg,Tang:2017qen,Ascencio-Sosa:2018lbk,Tang:2018rer,Coloma:2015kiu,Liao:2016orc,Masud:2017bcf,Kuchibhatla:2018grr,Deepthi:2017gxg}). It is a question of whether we are able to test different flavor and CP-violation models directly in running accelerator neutrino experiments like T2K~\cite{Abe:2011sj,Abe:2018wpn}and NO$\nu$A~\cite{Adamson:2016tbq,Adamson:2016xxw,NOvA:2018gge}, and future neutrino oscillation experiments like DUNE~\cite{Alion:2016uaj} and T2HK~\cite{Abe:2018uyc}. In this paper, we shall determine the potential of current and upcoming neutrino facilities to test the tri-direct CP approach, and the sensitivity regions of oscillation parameters will be presented.

The paper is organized as follows: we {firstly} review the tri-direct CP symmetry in Sec.~\ref{sec:review}. In Sec.~\ref{sec:exp}, we investigate the precision measurements of oscillation parameters, either represented by standard three neutrino mixing parameters or denoted by the benchmark model parameters, in running experiments such as T2K and NO$\nu$A. We expect better sensitivities in future neutrino experiments, such as T2HK, DUNE and Jiangmen Underground Neutrino Observatory (JUNO). {In Sec.~\ref{sec:result}, we show our simulation results. We study the precision of model parameters for different experimental configurations, and then discuss a degeneracy problem, which can be resolved by including JUNO data. We further study how to constrain oscillation parameters with the restriction of the tri-direct CP model, before we discuss two benchmark models.} Finally, we summarize our results in Sec.~\ref{sec:summary}.

\section{Review of Tri-direct CP symmetry models}\label{sec:review}
Let us firstly recapitulate a benchmark tri-direct CP model proposed in~\cite{Ding:2018fyz}. This model is based on the $S_4$ flavor symmetry and CP symmetry. The flavor group $S_4$ and CP are broken to the subgroups
$Z^T_3$, $Z^{TST^2}_2\times X_{\text{atm}}$ and $Z_2^U\times X^{\text{sol}}$ in the charged lepton, atmospheric neutrino and solar neutrino sectors, respectively, where $S$, $T$, $U$ are the generators of $S_4$ and $X_{\text{atm}}=SU$ and $X^{\text{sol}}=U$ denote the residual CP symmetry. In the generic tri-direct CP paradigm, the structure of the neutrino and charged lepton mass matrices essentially arise from the vacuum alignment of flavon fields which are fixed by the residual symmetry. In the working basis of~\cite{Ding:2018fyz}, the residual flavor symmetry $Z^T_3$ enforces that the charged lepton mass matrix is diagonal~\cite{Ding:2018fyz}. The atmospheric and solar flavon vacuum alignments are determined to be $\langle\phi_{\text{atm}}\rangle\propto\left(1, \omega^2, \omega\right)^T$ and $\langle\phi_{\text{sol}}\rangle\propto\left(1, x, x\right)^T$,
where $\omega=e^{2\pi i/3}$ is a cube root of unity and the parameter $x$ is real because of the imposed CP symmetry. As a result, the Dirac neutrino mass matrix reads as
\begin{equation}
m_{D}=\begin{pmatrix}
y_{a}      ~&~    y_{s}  \\
\omega y_{a}  ~&~  x y_{s} \\
\omega^2y_{a}  ~&~  x y_{s}
\end{pmatrix} \,.
\end{equation}
The right-handed neutrino Majorana mass matrix is diagonal
\begin{equation}
m_{N}=\begin{pmatrix}
M_{\textrm{atm}}  &  0  \\
0  &   M_{\textrm{sol}}
\end{pmatrix}\,.
\end{equation}
The light effective left-handed Majorana neutrino mass matrix is given by the seesaw formula: 
\begin{equation}
\label{eq:mnu}  m_{\nu}=m_{a}\begin{pmatrix}
 1 &~ \omega  &~ \omega ^2 \\
 \omega  &~ \omega ^2 &~ 1 \\
 \omega ^2 &~ 1 &~ \omega  \\
\end{pmatrix}+e^{i\eta}m_{s}
\begin{pmatrix}
 1 &~  x &~  x \\
 x &~ x^2 &~ x^2 \\
 x &~ x^2 &~ x^2 \\
\end{pmatrix}\,,
\end{equation}
where $m_a=|y^2_a/M_{\text{atm}}|$, $m_s=|y^2_s/M_{\text{sol}}|$, and the only physically important phase $\eta$ depends on the relative phase between $y^2_a/M_{\text{atm}}$ and $y^2_s/M_{\text{sol}}$. It is noteworthy that only four parameters $m_a$, $m_s$, $\eta$ and $x$ are involved to describe both neutrino masses and lepton mixing parameters. As a consequence, this model is quite predictive. The low-energy phenomenology of this model has been studied both numerically and analytically in~\cite{Ding:2018fyz}. {Although $x$ and the relative phase $\eta$ are free parameters in the general setup of tri-direct CP, they can be fixed to some particular values
through the vacuum alignment technique in a discrete flavor symmetry model. It is found that a quite good fit to the experimental data can be obtained
for certain choices of $x$ and $\eta$, two benchmark examples are $x=-7/2$, $\eta=\pi$ and $x=-4$, $\eta=5\pi/4$.
In these benchmark models, the corresponding vacuum alignments take a simple form such that they can be easily realized in
concrete models~\cite{Ding:2018fyz,Ding:2018tuj}. Moreover, the neutrino mass matrix as well as neutrino masses and mixing parameters
only depend on two free parameters $m_a$ and $m_s$ in the benchmark models and the experimental data can be accommodated very well.}

The neutrino mass spectrum is predicted to be normal ordering in this model, and the lightest neutrino is massless $m_1=0$. The other two non-vanishing neutrino masses $m_2$ and $m_3$ are expressed in terms of the input parameters as follows.
\begin{eqnarray}\label{eq:m2m3_TD}
\nonumber&&
m^2_2=\frac{1}{2}\left[|y|^2+|w|^2+2|z|^2-\sqrt{(|w|^2-|y|^2)^2+4|y^{*}z+wz^{*}|^2}\right], \\
&&
m^2_3=\frac{1}{2}\left[|y|^2+|w|^2+2|z|^2+\sqrt{(|w|^2-|y|^2)^2+4|y^{*}z+wz^{*}|^2}\right]\,,
\end{eqnarray}
where
\begin{eqnarray}
\nonumber && y=\frac{5x^2+2x+2}{2\left(x^2+x+1\right)}(m_{a}+e^{i \eta } m_{s})\,, \\
\nonumber && z=-\frac{\sqrt{5x^2+2x+2}}{2\left(x^2+x+1\right)}\left[ (x+2)m_{a}-x(2x+1)e^{i \eta }m_{s}\right]\,, \\
&& w=\frac{1}{2(x^2+x+1)}\left[(x+2)^2m_{a}+x^2\left(2x+1\right)^2e^{i \eta } m_{s}\right]\,.
\end{eqnarray}
As regards the predictions for lepton flavor mixing, the first column of the mixing matrix is determined to be proportional to $(\sqrt{3}\,x, \sqrt{x^2+x+1}, \sqrt{x^2+x+1})^{T}$, the other two columns are uniquely fixed by the input parameters, and the lepton mixing matrix is of the form~\cite{Ding:2018fyz}
\begin{eqnarray}
\label{eq:PMNS}U=\frac{1}{\sqrt{2}}\begin{pmatrix}
 \frac{ \sqrt{6} x}{\sqrt{5x^2+2x+2}} &~  2i\sqrt{\frac{x^2+x+1}{5x^2+2x+2}} \cos \theta  &~ 2i\sqrt{\frac{x^2+x+1}{5x^2+2x+2}} e^{i \psi }\sin \theta  \\
 \sqrt{\frac{2(x^2+x+1)}{5x^2+2x+2}} &~ -e^{-i \psi } \sin \theta -\frac{i \sqrt{3} x \cos\theta }{\sqrt{5x^2+2x+2}} &~ \cos \theta -\frac{i \sqrt{3}x e^{i \psi }  \sin \theta }{\sqrt{5x^2+2x+2}} \\
 \sqrt{\frac{2(x^2+x+1)}{5x^2+2x+2}} &~ e^{-i \psi } \sin \theta -\frac{i \sqrt{3}x\cos \theta }{\sqrt{5x^2+2x+2}} &~ -\cos \theta -\frac{i \sqrt{3}x e^{i \psi}\sin\theta}{\sqrt{5x^2+2x+2}} \\
\end{pmatrix}\,.
\end{eqnarray}
The angles $\theta$ and $\psi$ are specified by
\begin{eqnarray}
\nonumber&&\sin\psi=\frac{\Im\left(y^{*}z+wz^{*}\right)}{|y^{*}z+wz^{*}|},\quad \cos\psi=\frac{\Re\left(y^{*}z+wz^{*}\right)}{|y^{*}z+wz^{*}|}\,,\\
\nonumber&&\sin2\theta=\frac{2|y^{*}z+wz^{*}|}
{\sqrt{(|w|^2-|y|^2)^2+4|y^{*}z+wz^{*}|^2}},\\
&&\cos2\theta=\frac{|w|^2-|y|^2}{\sqrt{(|w|^2-|y|^2)^2+4|y^{*}z+wz^{*}|^2}}\,.
\end{eqnarray}
%Consequently we can extract the following results for the mixing angles,
As a consequence, we find the exact expressions for the mixing angles are
\begin{eqnarray}
\nonumber \sin^2\theta_{13}&=&\frac{2\left(x^2+x+1\right)\sin^2\theta}{5x^2+2x+2}\,,\\
\nonumber \sin^2\theta_{12}&=&1-\frac{3x^2 }{3x^2+2\left(x^2+x+1\right) \cos^2\theta }\,,\\
\label{eq:angles}\sin^2\theta_{23}&=&\frac{1}{2}+\frac{x\sqrt{3\left(5x^2+2x+2\right)}  \sin2\theta\sin\psi }{2\left[3x^2+2\left(x^2+x+1\right) \cos^ 2 \theta\right]}\,.
\end{eqnarray}
We see that the solar and reactor mixing angles satisfy the following sum rule
\begin{equation}\label{eq:correlation_mix_angles}
\cos^2\theta_{12}\cos^2\theta_{13}=\frac{3x^2}{5 x^2+2x+2}\,.
\end{equation}
Moreover, from the first column of the mixing matrix in Eq.~\eqref{eq:PMNS}, we can obtain a sum rule for $\cos\delta_\mathrm{CP}$ in terms of the lepton mixing angles
\begin{equation}
\cos\delta_\mathrm{CP}=\frac{ \cot 2 \theta_{23} \left[3x^2-\left(4x^2+ x+1\right)\cos^2\theta_{13}\right]}{\sqrt{3} \left|x\right| \sin \theta_{13} \sqrt{\left(5x^2+2x+2\right)\cos^2\theta_{13}-3x^2}}\,.
\end{equation}
If the atmospheric mixing angle is maximal, this sum rule implies that the Dirac CP phase would be maximal (i.e. $\delta_\mathrm{CP}=\pm\pi/2$) as well. Furthermore, the result for the Jarlskog invariant is
\begin{equation}
J_\mathrm{CP}=\frac{\sqrt{3} x\left(x^2+x+1\right)\sin2\theta\cos\psi}{2\left(5 x^2+2x+2\right)^{3/2}}\,,
\end{equation}
from which we can extract the value of $\sin\delta_\mathrm{CP}$,
\begin{equation}\label{eq:cosdelta}
\sin\delta_\mathrm{CP}= \pm\csc 2 \theta_{23} \sqrt{1+\frac{\left(x^2+x+1\right)^2 \cot ^2\theta_{13} \cos ^22 \theta_{23}}{3x^2 \left[3x^2 \tan ^2\theta_{13}-2 \left(x^2+x+1\right)\right]}}~\,,
\end{equation}
with ``$+$'' for $x\cos\psi>0$ and ``$-$'' for $x\cos\psi<0$. The above results for $\cos\delta_\mathrm{CP}$ and $\sin\delta_\mathrm{CP}$ allow us to fix the value of $\delta_\mathrm{CP}$. Comprehensive numerical analyses show that the allowed region of the parameters $x$, $\eta$, $r=m_a/m_s$ and $m_a$ are $-5.475\leq x\leq -3.370$, $0.455\pi\leq\eta\leq 1.545\pi$, $0.204\leq r\leq 0.606$ and $3.343~ \mathrm{meV}\leq m_a\leq 4.597~\mathrm{meV}$ respectively in order to accommodate the experimental data on neutrino masses and lepton mixing angles~\cite{Esteban:2016qun}. It is remarkable that both solar mixing angle and Dirac CP phase are predicted to lie in a narrow range $0.329\leq\sin^2\theta_{12}\leq0.346$ and $1.371\pi\leq\delta_\mathrm{CP}\leq1.629\pi$ in this model.

For the benchmark values of the vacuum parameters $x=-7/2$ and $\eta=\pi$, the effective light neutrino mass matrix in Eq.~\eqref{eq:mnu} {only depends} on two free parameters $m_a$ and $m_s$.
Using the general results presented above, we find that the lepton mixing matrix is of the form
\begin{equation}\label{modelA}
U=\frac{1}{5\sqrt{6}}\begin{pmatrix}
 7 \sqrt{2} ~&~ -2  \sqrt{13}\,i \cos \theta  ~&~ 2  \sqrt{13}  \sin \theta  \\
 \sqrt{26} ~&~ 7 i \cos \theta -5 \sqrt{3}  \sin \theta  ~&~ -7 \sin \theta +5\sqrt{3}\,i \cos \theta   \\
 \sqrt{26} ~&~ 7 i \cos \theta +5\sqrt{3}  \sin \theta   ~&~ -7\sin \theta -5 \sqrt{3}\,i \cos \theta
\end{pmatrix}\,,
\end{equation}
where
\begin{equation}
\sin2\theta=\frac{10 |14r-1|}{13 \sqrt{4+32r+289 r^2}}, \quad \cos2\theta=\frac{3 \left(8+57r\right)}{13 \sqrt{4+32r+289 r^2}}\,,
\end{equation}
with $r=m_s/m_a$. The lepton mixing angles read
\begin{equation}
\label{eq:angles_bm1}\sin^2\theta_{13}=\frac{26}{75}\sin^2\theta,\quad
\sin^2\theta_{12}=\frac{26\cos^2\theta}{62+13\cos2\theta },\quad
\sin^2\theta_{23}=\frac{1}{2}\,,
\end{equation}
and the Jarlskog invariant is
\begin{equation}
J_\mathrm{CP}=-\frac{91}{750\sqrt{3}} \sin2\theta \,,
\end{equation}
which implies that the Dirac CP phase is exactly maximal, i.e.
\begin{equation}
\delta_\mathrm{CP}=-\pi/2\,.
\end{equation}
Notice that both $\theta_{23}$ and $\delta_\mathrm{CP}$ are predicted to be maximal and they are favored by the latest data from T2K~\cite{Abe:2017uxa} and NO$\nu$A~\cite{Adamson:2017gxd,NOvA:2018gge}, the reason is the neutrino mass matrix of Eq.~\eqref{eq:mnu} fulfills the $\mu-\tau$ reflection symmetry in this case. The lightest neutrino masses as functions of $m_a$ and $r$ are
\begin{eqnarray}
\nonumber&&m^2_1=0\,,\\
\nonumber && m^2_2=\frac{9}{8} m^2_{a}\left(4-18r+289 r^2-\left|2-17r\right|\sqrt{4+32r+289r^2}  \right)\,, \\
&& m^2_3=\frac{9}{8} m^2_{a}\left(4-18r+289r^2+\left|2-17r\right|\sqrt{4+32r+289r^2}  \right)\,.
\end{eqnarray}
In order to accommodate the experimental values of the mixing angles and neutrino mass splittings $\Delta m^2_{21}$ and $\Delta m^2_{31}$~\cite{Esteban:2016qun}, we find that $m_a$ and $r$ are constrained to lie in rather narrows regions $3.560\,\mathrm{meV}\leq m_a\leq3.859\,\mathrm{meV}$ and $0.5282 \leq r\leq0.5904$. Accordingly the allowed regions of the reactor and solar mixing angles are strongly constrained $0.02206\leq\sin^2\theta_{13}\leq0.02349$ and $0.3310\leq\sin^2\theta_{12}\leq0.3319$.

Then we proceed to discuss the second representative values of the vacuum parameters $x=-4$ and $\eta=5\pi/4$, the lepton mixing matrix reads as
\begin{equation}\label{modelB}
U=\frac{1}{\sqrt{74}}\left(
\begin{array}{ccc}
4\sqrt{3} ~&~ -i \sqrt{26}\cos\theta ~&~ -i\sqrt{26}\, e^{i\psi} \sin\theta \\
\sqrt{13} ~&~ 2i\sqrt{6}\cos\theta-\sqrt{37}\, e^{-i\psi}\sin\theta  ~&~ 2i\sqrt{6}\, e^{i\psi}\sin\theta+\sqrt{37}\cos\theta \\
\sqrt{13} ~&~ 2i\sqrt{6}\cos\theta+\sqrt{37}\,e^{-i\psi}\sin\theta  ~&~ 2i\sqrt{6}\,e^{i\psi}\sin\theta-\sqrt{37}\cos\theta
\end{array}
\right)\,,
\end{equation}
where a Majorana phase matrix is omitted, and the parameters $\theta$ and $\psi$ are functions of the mass ratio $r$,
\begin{eqnarray}
\nonumber&&\tan2\theta=\frac{2\sqrt{37}\sqrt{4225 r^2+9 \left(\sqrt{2}-25 r+154\sqrt{2}\,r^2\right)^2}}{15\left(-7+3\sqrt{2}\,r+781r^2\right)}\,,\\
&&\tan\psi=-\frac{65r}{3\left(\sqrt{2}-25r+154\sqrt{2}\,r^2\right)}\,.
\end{eqnarray}
The expressions of the mixing angles are
\begin{equation}
\label{eq:angles_bm2}\sin^2\theta_{13}=\frac{13}{37}\sin^2\theta,\quad
\sin^2\theta_{12}=\frac{26\cos^2\theta}{61+13\cos2\theta },\quad
\sin^2\theta_{23}=\frac{1}{2}-\frac{2\sqrt{222}\sin2\theta\sin\psi}{61+13\cos2\theta}\,.
\end{equation}
The Jarlskog CP invariant takes the form
\begin{equation}
J_{CP}=-\frac{13}{74}\sqrt{\frac{6}{37}}\sin2\theta\cos\psi\,.
\end{equation}
The sum rules for the Dirac CP phase in terms of lepton mixing angles are given by
\begin{eqnarray}
\nonumber&&\cos\delta_\mathrm{CP}=\frac{(35-61\cos2\theta_{13})\cot2\theta_{23}}{8\sin\theta_{13}\sqrt{111\cos2\theta_{13}-33}}\,,\\
&&\sin\delta_\mathrm{CP}=-\csc2\theta_{23}\sqrt{1-\frac{169\cot^2\theta_{13}\cos^22\theta_{23}}{96(13-24\tan^2\theta_{13})}}\,.
\end{eqnarray}
It noteworthy that all the lepton mixing angles as well as $\delta_\mathrm{CP}$ only depend on the parameter $r$ through $\theta$ and $\psi$ in this case. Moreover, the results for the light neutrino masses are
\begin{small}
\begin{eqnarray}
\nonumber&&m^2_1=0\,,\\
\nonumber&&m^2_2=\frac{1}{2}m^2_a\left(9-25\sqrt{2}\,r+1089r^2-\sqrt{81-450\sqrt{2}(1+121r^2)r+[(1089r)^2-1052]r^2}\right)\,,\\
&&m^2_3=\frac{1}{2}m^2_a\left(9-25\sqrt{2}\,r+1089r^2+\sqrt{81-450\sqrt{2}(1+121r^2)r+[(1089r)^2-1052]r^2}\right)\,.
\end{eqnarray}
\end{small}
In order to describe the experimentally measured values of both lepton mixing angles and neutrino mass squared differences, we find the allowed ranges of the input parameters are $3.568\,\mathrm{meV}\leq m_a\leq 3.871\,\mathrm{meV}$ and $0.3983\leq r\leq0.4473$. As a consequence, the solar and reactor mixing angles are constrained to lie in the narrow intervals $0.02254\leq\sin^2\theta_{13}\leq0.02280$ and $0.3362\leq\sin^2\theta_{12}\leq0.3364$, and the atmospheric mixing angle is predicted to be in the second octant $0.5559\leq\sin^2\theta_{23}\leq0.5636$. The predicted values of $\delta_\mathrm{CP}$ are distributed around $3\pi/2$, namely $1.582\pi\leq\delta_\mathrm{CP}\leq1.594\pi$.

%%%%%%%%%%%%%%%%%%%%%%%%%%%%%%%%%%%%%%%%%%%%%%%%%%%%%%%%%%%%%%%

\begin{table}[t!]
\renewcommand{\tabcolsep}{0.7mm}
\renewcommand{\arraystretch}{1.3}
\small
\centering
\begin{tabular}{|c|c| c| c| c | c| c| c| c| c| c|c |c|c|c|}  \hline \hline
 $x$   & $\eta$  & $m_{a}(\text{meV})$ & $r$ 	 & $\chi^2_{\text{min}}$ &  $\sin^2\theta_{13}$  &$\sin^2\theta_{12}$  & $\sin^2\theta_{23}$  & $\delta_{CP}/\pi$ &  $\beta/\pi$ & $m_2(\text{meV})$ & $m_3(\text{meV})$ & $m_{ee}(\text{meV})$ \\   \hline

$-\frac{7}{2}$ &   $\pi$ &   $3 .716$ &  $0 .557$ & $17 .524$ & $0 .0227$ &  $0 .331$ &  $0 .5$ &  $-0.5$ &  $0$ &  $8 .611$ & $50 .232$ &  $1 .647$ \\ \hline

$-4$ &  $\frac{5\pi }{4}$ &  $3 .723$ &  $0 .421$ & $5 .168$ & $0 .0226$ &  $0 .336$ &  $0 .560$ &  $-0.412$ &  $0 .264$ & $8 .603$ & $50 .242$ &  $2 .840$  \\  \hline \hline
\end{tabular}
\caption{\label{tab:bf_N4}The best fit values of the lepton mixing angles, CP violation phases, neutrino masses and the effective Majorana mass $m_{ee}$ for the benchmark values $(x, \eta)=(-7/2, \pi), (-4, 5\pi/4)$ of the tri-direct CP model.}
\end{table}

%%%%%%%%%%%%%%%%%%%%%%%%%%%%%%%%%%%%%%%%%%%%%%%%%%%%%%%%%%%%%%%

Since the model is very predictive and the mixing angles as well as Dirac CP phase are constrained to lie in rather narrow regions, in particular we have $0.329\leq\sin^2\theta_{12}\leq0.346$ for the most general case, we expect the benchmark tri-direct model could be excluded in future neutrino experiments. If $\theta_{23}$ and $\delta_{CP}$ are measured precisely enough, the two values $x=-7/2, \eta=\pi$ and $x=-4, \eta=5\pi/4$ may be distinguished from each other. It will be nice to probe these features in detailed simulations of current and future neutrino oscillation experiments.

\section{Implementation of neutrino experiments in simulation}
\label{sec:exp}

In this section, we will introduce the current and future experiments -- T2K, NO$\nu$A, T2HK, DUNE and JUNO. All sensitivities in experiments are simulated in a state-of-the-art tool GLoBES~\cite{Huber:2004ka,Huber:2007ji} where the experimental details can be very nicely implemented by {an} Abstract Experimental Design Language (AEDL) file. As soon as the publicly available signal and background spectra are reproduced, we can safely claim the expected sensitivities in the precision measurements. In the simulation, input values of neutrino mixing parameters are taken as the best fit values of the latest NuFit4.0~\cite{Esteban:2018azc}:
$\sin^2\theta_{12}=0.310$, $\sin^2\theta_{13}=0.0224$,
$\sin^2\theta_{23}=0.580$, $\delta_{\mathrm{CP}}=215^\circ$, $\Delta m_{21}^2=7.39\times10^{-5}$ eV$^2$, $\Delta m_{31}^2=2.525\times10^{-3}$ eV$^2$.
In the current study, we will choose a normal mass hierarchy as a demonstration. In the meantime, the Preliminary Reference Earth Model (PREM) density profile is considered in the numerical calculations~\cite{prem:1981}.
We are using two methods to present our results:
\begin{itemize}
 \item {Standard three neutrino oscillations expressed by $\theta_{12}$, $\theta_{13}$, $\theta_{23}$, $\delta_\mathrm{CP}$, $\Delta m_{21}^2$ and $\Delta m_{31}^2$ are taken as the truth in Nature, we expect that precision measurements of mixing parameters are correlated{, and uncertainties of current global fit results are taken into account.} For given oscillation parameters, we define a set of parameters:
 \begin{equation}
  \overrightarrow{\mathcal{O}}=\{\theta_{12}\,,\theta_{13}\,,\theta_{23}\,,\delta_\mathrm{CP}\,,\Delta m_{21}^2\,, \Delta m_{31}^2 \}
 \end{equation}
and predict the expected {event rate} in {the bin $i$} $\mu_i(\overrightarrow{\mathcal{O}})$. We suppose a given experiment reconstructs neutrino spectra in $N$ bins sequentially. {The event rate} in { the bin $i$} is recorded as $n_i$. We can build a $\chi^2(\overrightarrow{\mathcal{O}})$ to quantify the sensitivity:
\begin{equation}\label{eq:chi_PMNS}
 \chi^2(\overrightarrow{\mathcal{O}})=\sum_{i=1}^N  \frac{\left[\mu_i(\overrightarrow{\mathcal{O}})-n_i\right]^2}{\sigma_i^2} \,.
\end{equation}
The final results come from a minimization of the summation of {the} $\chi^2(\overrightarrow{\mathcal{O}})$ in {every oscillation channel of all} experiments over a set of parameters, or the so-called marginalization.
 }
 \item {{Once we fit the model parameters}, the number of degrees of freedom is reduced from six to four, as shown in the previous section. We consider the following parameters from the tri-direct CP symmetry models: $x$, $\eta$, $m_{a}$ and $r$. In this case, we have to change the oscillation parameters predicted by the specific model:
\begin{equation}
  \overrightarrow{\mathcal{M}}=\{x\,,\eta\,,m_{a}\,,r \}
 \end{equation}
Other steps in the likelihood analysis will follow the same strategy as the above method, but replace the equation Eq.~(\ref{eq:chi_PMNS}) by
\begin{equation}\label{eq:chi_TD}
\chi^2(\overrightarrow{\mathcal{M}})=\sum_{i=1}^N  \frac{\left[\mu_i(\overrightarrow{\mathcal{O}}(\overrightarrow{\mathcal{M}}))-n_i\right]^2}{\sigma_i^2} \,.
\end{equation}
with the PMNS parameters as {functions} of model parameters $\overrightarrow{\mathcal{O}}(\overrightarrow{\mathcal{M}})$.
We can expect better measurements of input parameters after a combination of experimental results and symmetry-induced constraints from the theory.}
\end{itemize}

\subsection{T2K}
T2K stands for Tokai to Kamioka a long-baseline experiment in Japan.
In Tokai, muon neutrinos or antineutrinos are produced by bombarding a 30 GeV proton beam onto a graphite target station in the J-PARC accelerator center. The neutrino beams are detected {firstly} at the near detectors which are 280 meters away from the target station. {The far detector which reconstructs oscillated neutrino/antineutrino signals is Super-Kamiokande which has a fiducial mass of 22.5 kilotons and is 295 kilometers away with an off-axis angle of 2.5$^\circ$ from the beam direction.}
With the carefully chosen off-axis angle, the neutrino beam energy is peaked at about 0.6 GeV and matches the first maximum in the neutrino oscillation channels: P($\nu_{\mu}\to\nu_e$) and P($\bar{\nu}_\mu\to\bar{\nu}_e$).

In 2011, the T2K collaboration {published} their first result on P($\nu_{\mu}\to\nu_e$) with $1.43\times10^{20}$ Protons On Target (POT). It is the first hint of non-zero $\theta_{13}$ at $2.5\sigma$ confidence level (C.L.)~\cite{Abe:2011sj}.
In 2012, they {presented} an analysis of neutrino oscillation for P($\nu_{\mu}\to\nu_\mu$) based on the same POT data, where we have best-fit values of $\Delta m_{32}^2=2.63\times10^{-3}$ eV$^2$ and $\sin^22\theta_{23}=0.98$ in the three-neutrino mixing framework~\cite{Abe:2015ibe}.
In 2016, the first antineutrino result was published based on $4.01\times10^{20}$ POT, where we have best-fit values of $\Delta m_{32}^2=2.51\times10^{-3}$ eV$^2$ and $\sin^2\theta_{23}=0.45$.
The latest results of searching for CP violation in neutrino and antineutrino oscillations by T2K are based on $2.2\times10^{21}$ POT~\cite{Abe:2018wpn}.
In our simulation, we equally split $7.8\times10^{21}$ POT into two modes for T2K as the final total POT number.

\subsection{NO$\nu$A}

NO$\nu$A is a long-baseline neutrino oscillation experiment in the United States. Muon neutrinos or antineutrinos are produced by the NuMI beam at Fermilab. The experiment also adopts an off-axis angle of $14.6$ mrad to reach the first neutrino oscillation maximum at a peak energy of $2$ GeV, since the far detector using $14$ kt active scintillator is $810$ km away from the target station. The far detector is on the surface. An identical detector with a mass of $290$ ton scintillator is $100$ meter deep at a distance of $1$ km in order to monitor the neutrino flux and cancel the systematic uncertainties.

In 2016, the NO$\nu$A collaboration {published} their first result in the $\nu_{\mu}\to\nu_e$ channel~\cite{Adamson:2016tbq} and in the $\nu_{\mu}\to\nu_\mu$ channel~\cite{Adamson:2016xxw} based on $2.74\times10^{20}$ POT.
In 2017, they {updated} results on the electron neutrino appearance {channel} based on $6.05\times10^{20}$ POT~\cite{Adamson:2017qqn}. The degeneracy of $\theta_{23}$ shows up at $2.6\sigma$ C.L.
The latest results in a combination of neutrino and antineutrino runs are given in Ref.~\cite{NOvA:2018gge}.
In our simulation, we assume total $36\times 10^{20}$ POT for $\nu$ and $\bar{\nu}$ modes until $2024$ for NO$\nu$A.

\subsection{T2HK}
{An evolution of Water Cherenkov detectors from Kamiokande to Hyper-Kamiokande} makes it possible to conduct an upgrade of T2K to T2HK~\cite{Abe:2018uyc}. The HyperK detector will have 560 kt fiducial mass to reconstruct neutrino oscillation spectra. T2HK shares the same baseline of 295 km as T2K while the offaxis beam remains in the same direction with an upgraded proton beam at 1.3 MW. We assume T2HK is running in {the} neutrino mode in 2.5 years and in the antineutrino mode in 7.5 years. The second far detector in Korea is actively under {consideration}. In our simulation, we will keep the conservative option without the second far detector.

\subsection{DUNE}
DUNE is the next-generation accelerator neutrino oscillation experiment with a baseline of $1300$ km from FNAL to the underground laboratory in South Dakota. The experiment will search for CP violation in the leptonic sector and conduct precision measurements using appearance and disappearance channels by { $\nu_\mu$ and $\bar{\nu}_\mu$ beams}. DUNE is going to reconstruct oscillated neutrino spectra with a detector complex of four $10$-kt Liquid Argon Time Projection Chamber (LArTPC).  We adopt an AEDL file provided by Ref.~\cite{Alion:2016uaj}. We assume the experiment {will be running} in the neutrino/antineutrino mode in $3.5$ years, and adopt the 3-horn optimised {beam} design, which consists of the $62.5$ GeV proton beam with a power of $1.83\times10^{21}$ POT per year \cite{Papadimitriou:2017ytl,Tariq:2016ysu}.

\subsection{JUNO}

JUNO is a multi-purpose underground neutrino experiment, which will build a 20 kt liquid scintillator detector in South China and {is planned} to be online in 2021~\cite{Djurcic:2015vqa}.
The primary goal of JUNO~\cite{An:2015jdp} is to determine the neutrino mass {ordering} and precision measurement of oscillation parameters using the
reactor electron neutrino disappearance channel thanks to the unprecedented energy resolution of $3\%/\sqrt{E}$.
Regarding the precision {measurement} of $|\Delta m^{2}_{31}|$, $\Delta m^{2}_{21}$ and $\sin^{2}\theta_{12}$,
JUNO can reach the levels of 0.44\%, 0.59\% and 0.67\% respectively, after six years of data taking.
Moreover, the determination of the neutrino mass {ordering} at reactors is free from the contamination of matter effects~\cite{Li:2016txk} and possible new physics~\cite{Li:2018jgd}. JUNO will be rather robust when combined with accelerator neutrino experiments.
The sub-percent level precision for three of all six standard oscillation parameters will {certainly be powerful for selecting} the flavor-symmetry models.
In our simulation, we use the standard precision levels as our input priors to combine with accelerator neutrino experiments using
the state-of-the-art GLoBES tool.

\section{Model testing with neutrino oscillation experiments}\label{sec:result}

{In this section, we show our simulation results with the experiments introduced in Sec.~\ref{sec:exp}. The configurations we considered are the synergy of T2K and NO$\nu$A, DUNE, T2HK, the combination of all LBLs, and the interplay of LBLs and the reactor experiment JUNO. In Sec.~\ref{sec:precision}, we will firstly investigate the precision of four model parameters -- $x$, $\eta$, $r$ and $m_a$. One will see in those results, there is a degeneracy problem. In Sec.~\ref{sec:degeneracy}, we will explain how this degeneracy problem appears, and propose a way to resolve this problem. In the following section, we will study how the uncertainties are changed for oscillation parameters by tri-direct CP models. The above subsections are based on the general tri-direct CP model. In Tab.~\ref{tab:bf_N4}, we show two benchmark models. Finally, we will predict how these two benchmark models can be tested in future experiments.}

\subsection{Precision measurement of model parameters}\label{sec:precision}
\begin{figure}[!t]
 \includegraphics[width=1.\textwidth]{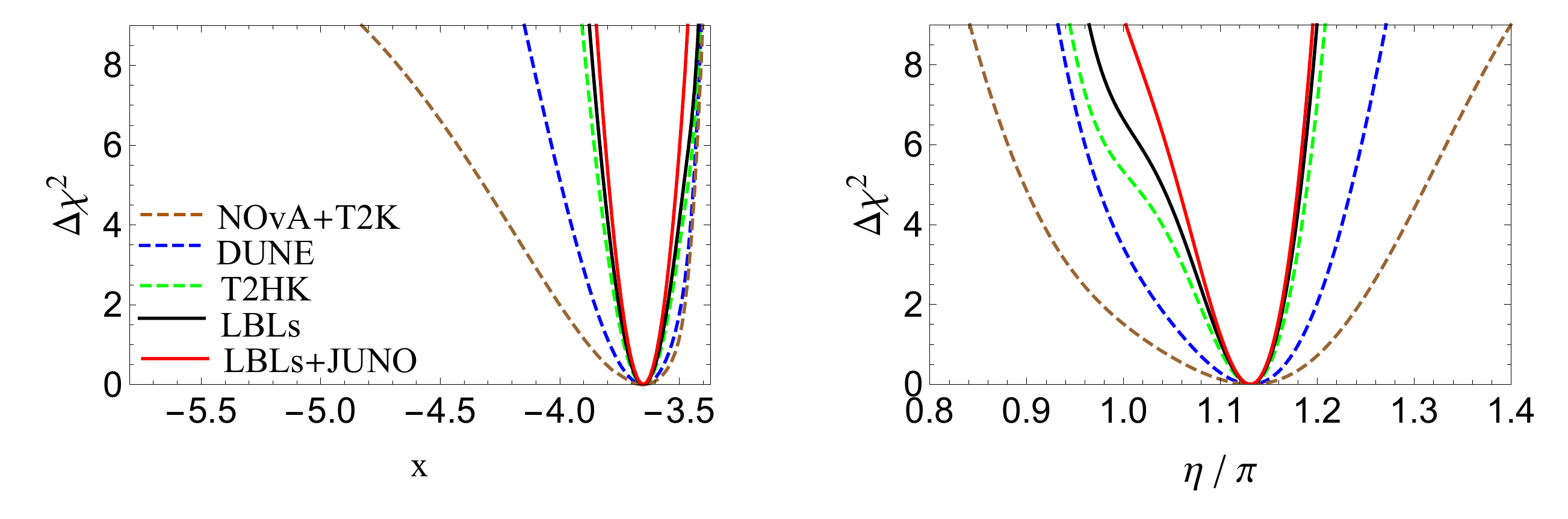}\\ \vspace{2mm}
  \includegraphics[width=1.\textwidth]{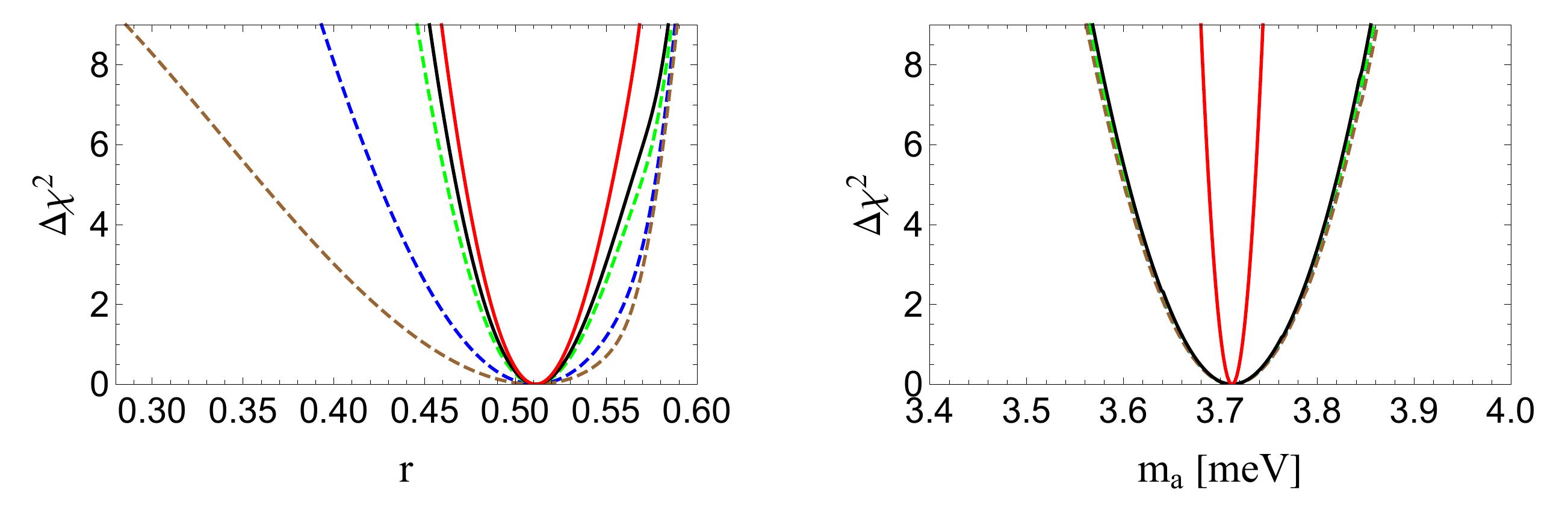}
\caption{\label{fig:model1D}The $\Delta\chi^2$ value of $x$, $\eta$, $r$, and $m_a$ in the framework of three-neutrino oscillations taking uncertainties of the NuFit4.0 results. True values for the model parameters are used  $(x,~\eta,~r,~m_a)=(-3.65,~1.1\pi,~0.5,~3.7\text{meV})$. The {experimental} configurations we considered are the combination of NO$\nu$A and T2K (dashed-brown), DUNE (dashed-blue), T2HK (dashed-green), the combination of all LBLs (solid-black), including all LBLs and JUNO (solid-red).}
\end{figure}

\begin{figure}[!t]
 \flushleft
\includegraphics[width=0.3\textwidth]{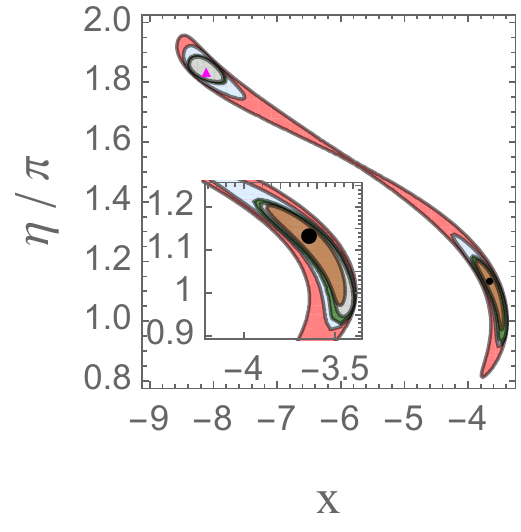} \hspace{1.25cm}
 \includegraphics[width=0.11\textwidth]{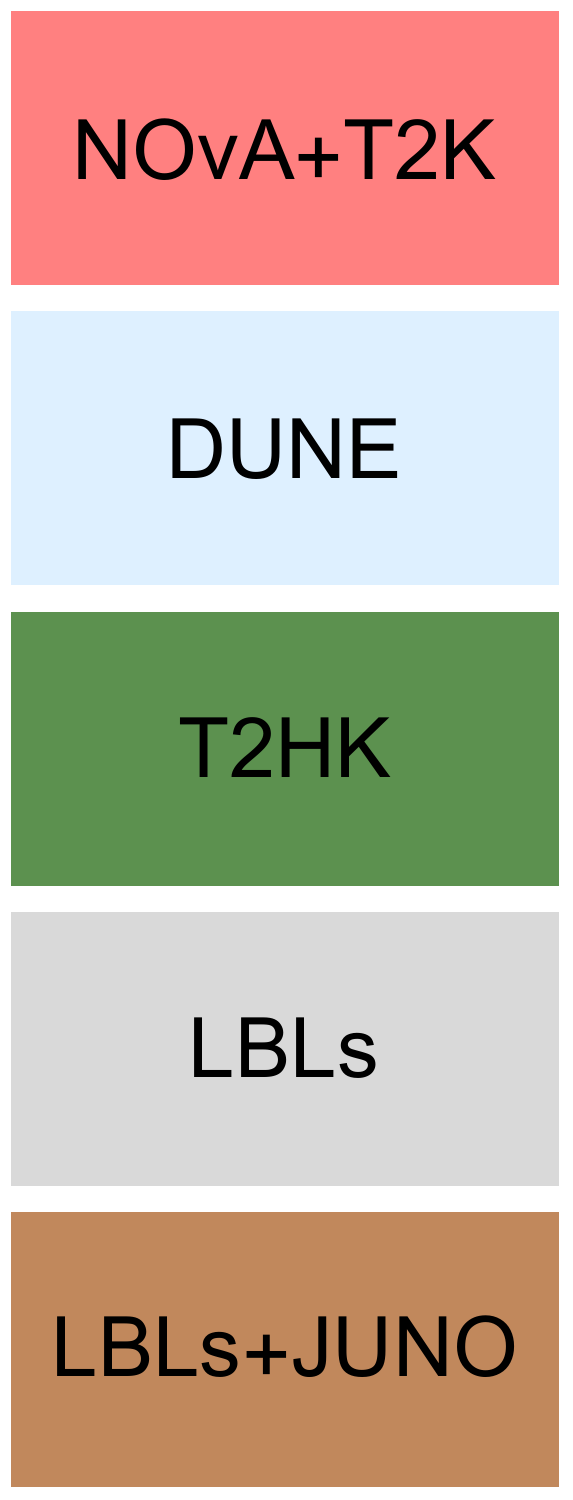}$~~~~~~$\\ \vspace{1mm}
 \includegraphics[width=0.3\textwidth]{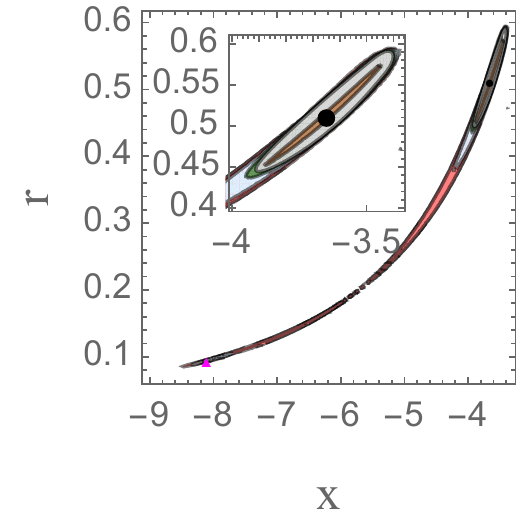}
 \includegraphics[width=0.3\textwidth]{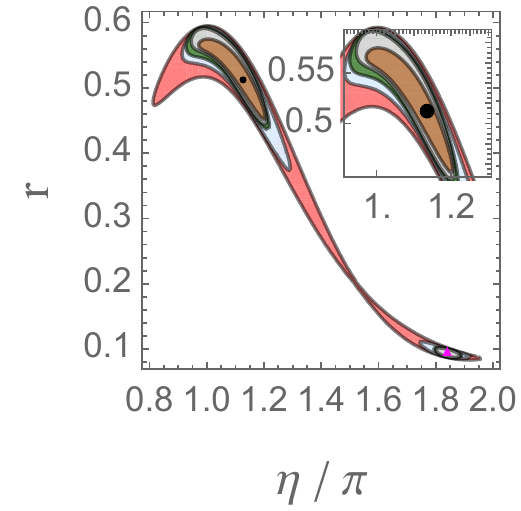}$~~~~~~$\\ \vspace{1mm}
 \includegraphics[width=0.3\textwidth]{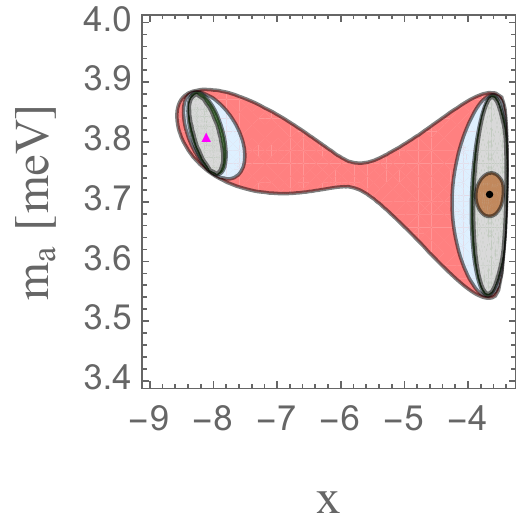}
 \includegraphics[width=0.3\textwidth]{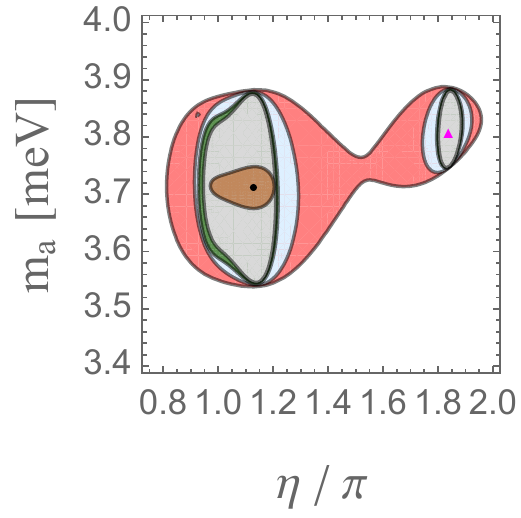}
 \includegraphics[width=0.3\textwidth]{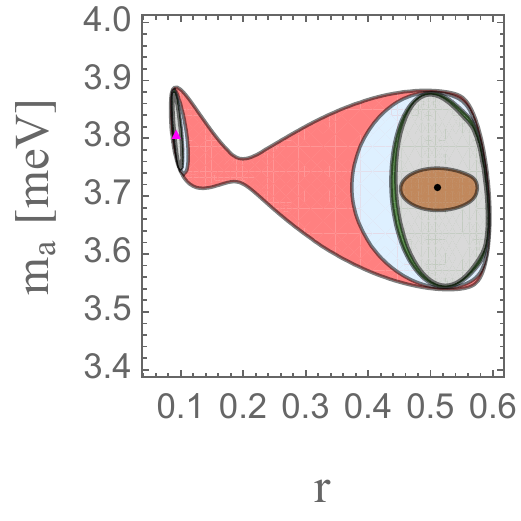}
 \caption{\label{fig:model2D}Precision measurements of any two model parameters at the 3$\sigma$ confidence level in the framework of three-neutrino oscillations taking uncertainties of the NuFit4.0 results. True values for the model parameters are used $(x,~\eta,~r,~m_a)=(-3.65,~1.1\pi,~0.5,~3.7\text{meV})$. We present the expected results from DUNE (light blue), T2HK (green), the combination NO$\nu$A and T2K (pink) and the synergy of all LBLs (light grey) and the interplay of LBLs and JUNO (brown). The black dot denotes the best fit values, while the magenta triangle is for the local minimum, where $r\sim0.1$, $\eta\sim1.84\pi$, $x\sim-8$ and $m_a\sim3.81$meV.}
\end{figure}

In Fig.~\ref{fig:model1D}, we show the $\Delta\chi^2$ values for each model parameter. We use the true values for the model parameters; $(x,~\eta,~r,~m_a)=(-3.65,~1.1\pi,~0.5,~3.7\text{meV})$, which is the best fit for the NuFit4.0 result. We also include the prior according to the NuFit4.0 result. We consider the configurations: DUNE (dashed-blue), T2HK (dashed-green), the combination of NO$\nu$A  and T2K (dashed-brown), the synergy {of} all LBLs (dashed-grey), and the optimised configuration by combing all LBLs and the JUNO experiment (solid-red). Except for the $m_a$ result, we see {a great improvement in the DUNE result compared to the T2K-and-NO$\nu$A combination.} T2HK further improves the measurements, and its performance is similar to the combination of all LBLs. This demonstrates the fact that T2HK dominates the contributions. The feature that the performance of T2HK is better than that of DUNE {reflects} the well-known result that T2HK works better than DUNE with fixed mass ordering, which is naturally imposed by the {tri-direct CP model}.
In more detail, for $x$ (the upper-left panel) the $3\sigma$ uncertainty {improves} from the T2K-and-NO$\nu$A combination ($\sim[-4.8,~-3.5]$) to DUNE ($\sim[-4.2,~-3.5]$) and T2HK ($\sim[-3.8,~-3.5]$). The combination of all LBLs performs similarly toT2HK.

Features and tendencies of each $\Delta\chi^2$ curves against $r$ (the lower-left panel) are similar to the result for $x$. The uncertainties at $3\sigma$ for the T2K-and-NO$\nu$A combination, DUNE, T2HK are $\sim[0.3,~0.6]$, $\sim[0.4,~0.6]$, $\sim[0.45,~0.6]${, respectively.} The $3\sigma$ uncertainty for combining all LBLs is almost the same as that for T2HK.
%
%The asymmetry are seen in both results for $x$ and $r$, and is caused by the existing data. 
The relative symmetry is seen in the result for $\eta$. The size of the $3\sigma$ uncertainty for the T2K-and-NO$\nu$A combination, DUNE and T2HK are about $0.3\pi$, $0.2\pi$, $0.15\pi${, respectively.} The {correlation between} $\eta$ and $r$ worsens the sensitivity for $\eta$ smaller than the assumed true value. Thanks to the high precision {of} T2HK and combing all LBLs, the degeneracy problem can be resolved when $\eta$ is very close to the true value. Therefore, we see a twist around $\eta=\pi$ for these two configurations. Details about this degeneracy will be introduced in Sec.~\ref{sec:degeneracy}.

For the above three parameters $x$, $\eta$ and $r$, it is hard to see the improvement by including the data of JUNO to those of all LBLs.
Data from JUNO is important for the $m_a$ measurement.
We see the overlapping of all curves for all LBL configurations (dashed-blue, dashed-green, dashed-brown, black curves) in the $m_a$ result. The uncertainty is mainly contributed from $\Delta m^2_{21}$, which is not measured well by LBLs. As a result, we see a great improvement by including data from JUNO, which well {measures} $\Delta m^2_{21}$.
%
%[text for the results with reactors]

We also show the $3\sigma$ ($\Delta\chi^2=11.83$) contour between {any two model parameters} in Fig.~\ref{fig:model2D}.
We see some correlations among $x$, $\eta$ and $r$ for all configurations on $x-\eta$, $x-r$ and $\eta-r$ planes. This correlation is consistent with what we see in Eq.~(\ref{eq:mnu}), in which $m_a$ is less dependent on the other three parameters. We discover a degeneracy problem related to this correlation for all possible LBL configurations-- the combination of NO$\nu$A and T2K, DUNE, T2HK, and all-LBL synergy. This degeneracy is mainly caused by the poor measurement of $\theta_{12}$. More details about this degeneracy can be seen in Sec.~\ref{sec:degeneracy}. These correlations are not removed even if we include JUNO, but combing LBLs and JUNO data can {resolve} the degeneracy problem.

\subsection{Breaking degeneracies}\label{sec:degeneracy}
\label{app:degeneracy}
\begin{figure}[!t]
 \includegraphics[width=0.3\textwidth]{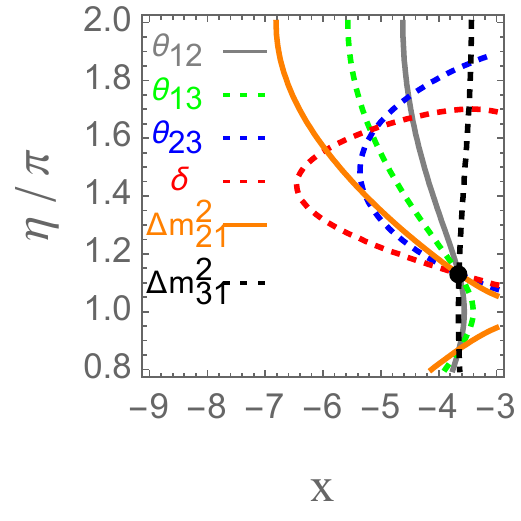}
 \includegraphics[width=0.3\textwidth]{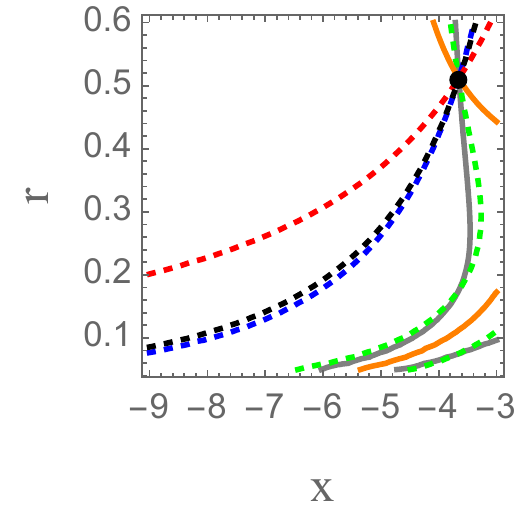}
 \includegraphics[width=0.3\textwidth]{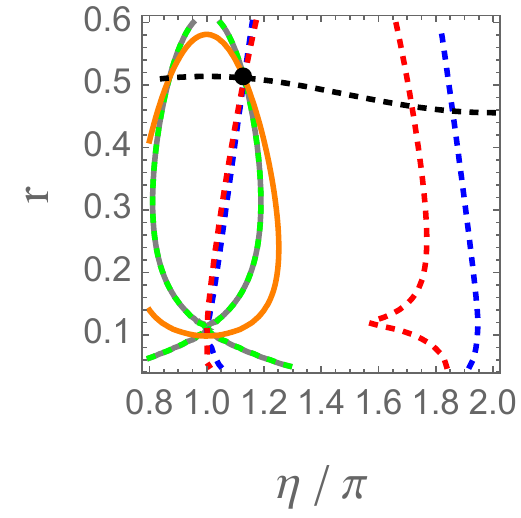}\\ \vspace{5mm}
 \includegraphics[width=0.3\textwidth]{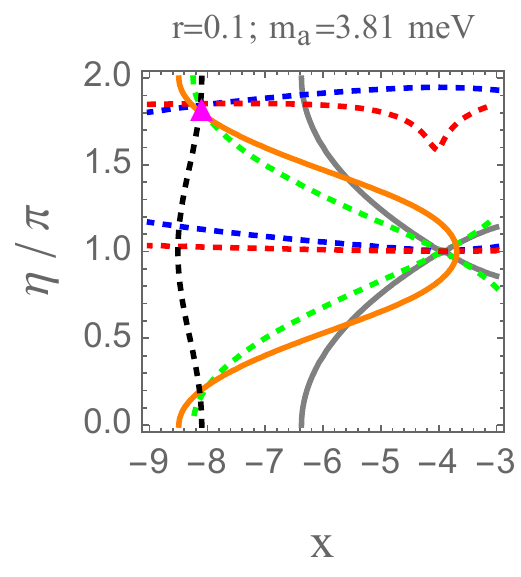}
 \includegraphics[width=0.3\textwidth]{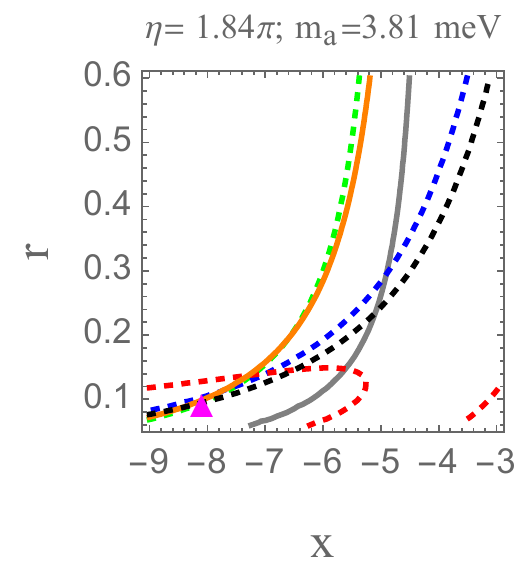}
 \includegraphics[width=0.3\textwidth]{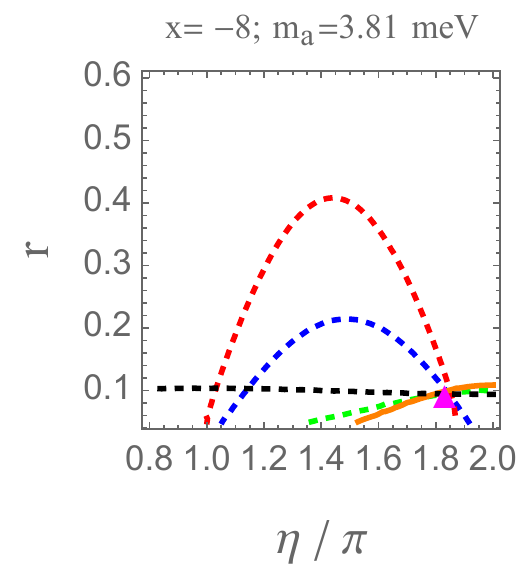}
 \caption{\label{fig:deg}The contours for $\theta_{12}\sim35.3^\circ$ (grey), $\theta_{13}\sim8.6^\circ$ (short-dashed-green), $\theta_{23}\sim47^\circ$ (short-dashed-blue), $\delta\sim279^\circ$ (short-dashed-red), $\Delta m_{21}^2\sim 7.4\times 10^{-5}$ eV$^2$ (short-dashed-orange), and $\Delta m_{31}^2\sim 2.52\times 10^{-3}$ eV$^2$ (black) on {the} $x-\eta$, $x-r$, and $\eta-r$ planes. In the upper panels, we let model parameters {be} the best fit values, except for those {which are varied}. {In the lower panels} we focus on the degeneracy regions, where $r\sim0.1$, $\eta\sim1.84\pi$, $x\sim-8$ and $m_a\sim3.81$meV. The gray curve for $\theta_{12}$ is below $r=0.07$ so that it is hardly visible. We show the true values and the local minimum {using} black dots and magenta triangles respectively.
}
\end{figure}

The degeneracy in Fig.~\ref{fig:model2D} can be understood by the equal-oscillation-parameter-value contours on different planes as shown in Fig.~\ref{fig:deg}. In Fig.~\ref{fig:deg}, we show these contours on {the} $x-\eta$, $x-r$, and $\eta-r$ planes. In the upper panels, we set the model paramters at the true values $(x,~\eta,~r,~m_a)=(-3.65,~1.1\pi,~0.5,~3.7~\text{meV})$, which predicts the value for oscillation parameters  $\theta_{12}\sim35.3^\circ$ (grey), $\theta_{13}\sim8.6^\circ$ (short-dashed-green), $\theta_{23}\sim47^\circ$ (short-dashed-blue), $\delta\sim279^\circ$ (short-dashed-red), $\Delta m_{21}^2\sim 7.4\times 10^{-5}$ eV$^2$ (orange), and $\Delta m_{31}^2\sim 2.52\times 10^{-3}$ eV$^2$ (short-dashed-black). The contours are shown with these conditions. Therefore, the intersection of all contours is at the assumed true values. {In} the lower panels, we focus on the degeneracy region: for the left, middle, and the right panels, we set $r\sim0.1$, $\eta\sim1.84\pi$, $x\sim-8$ and $m_a\sim3.81$meV. We see that the local minimum of the degeneracy region (magenta triangles) takes place where the green, blue, red, orange and black curves meet together or go very close. LBL experiments are not sensitive to $\theta_{12}$ (grey curve) and $\Delta m_{21}^2$ (orange curve). 
As a result, these LBL experiments cannot exclude this region by improving precision. This also explains {why once} we include reactor data that is sensitive to $\theta_{12}$, the degeneracy region is excluded.
{One may notice that the different curves do not intersect at the magenta triangle in the $x-r$ plane and there is the gray curve of $\theta_{12}$ in the last panel for $\eta-r$. The reason is that the triangle presents a local minimum at which $11.83>\Delta\chi^2>9$. Though that is a local minimum, it does not need to cross all curves. The grey curve of $\theta_{12}$ is below r=0.07, where the bottom of the panel is. We did not show it in this panel because the main feature here crossing or going close to $\theta_{13}$, $\theta_{23}$, $\delta$ and $\Delta m^2_{31}$ results in the degeneracy issue for LBLs.
}

\subsection{Standard oscillation parameters under tri-direct CP symmetry model}
\begin{figure}[!t]
 \includegraphics[width=1.\textwidth]{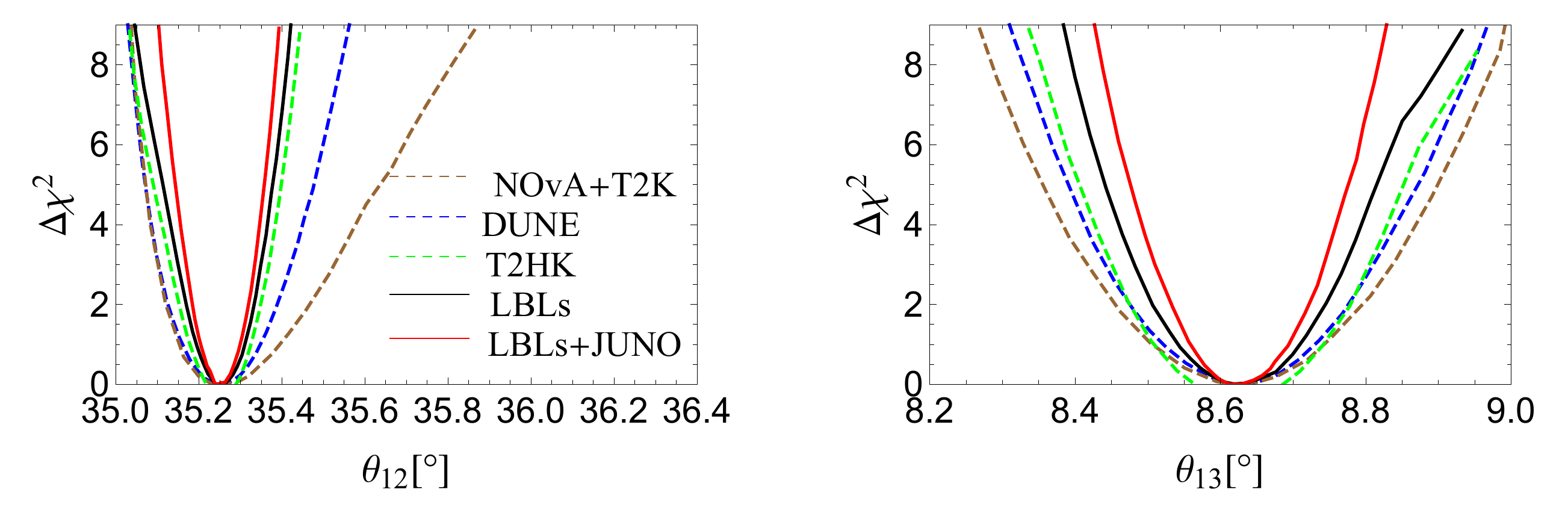}\\ \vspace{2mm}
 \includegraphics[width=1.\textwidth]{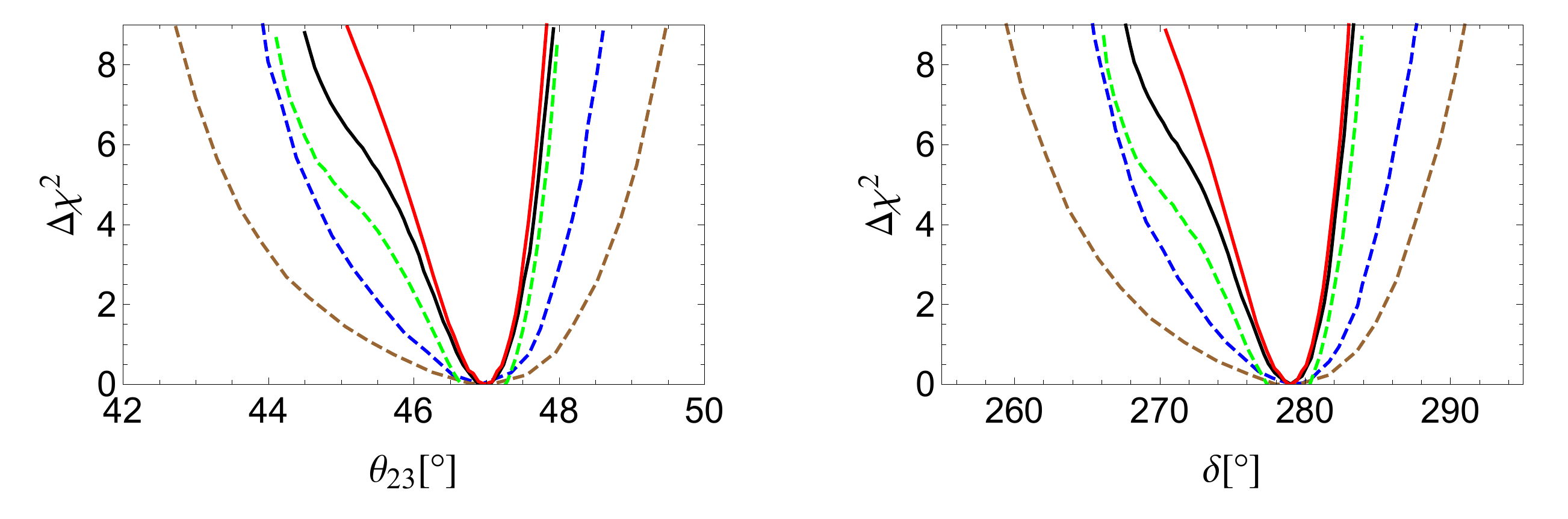}\\ \vspace{2mm}
  \includegraphics[width=1.\textwidth]{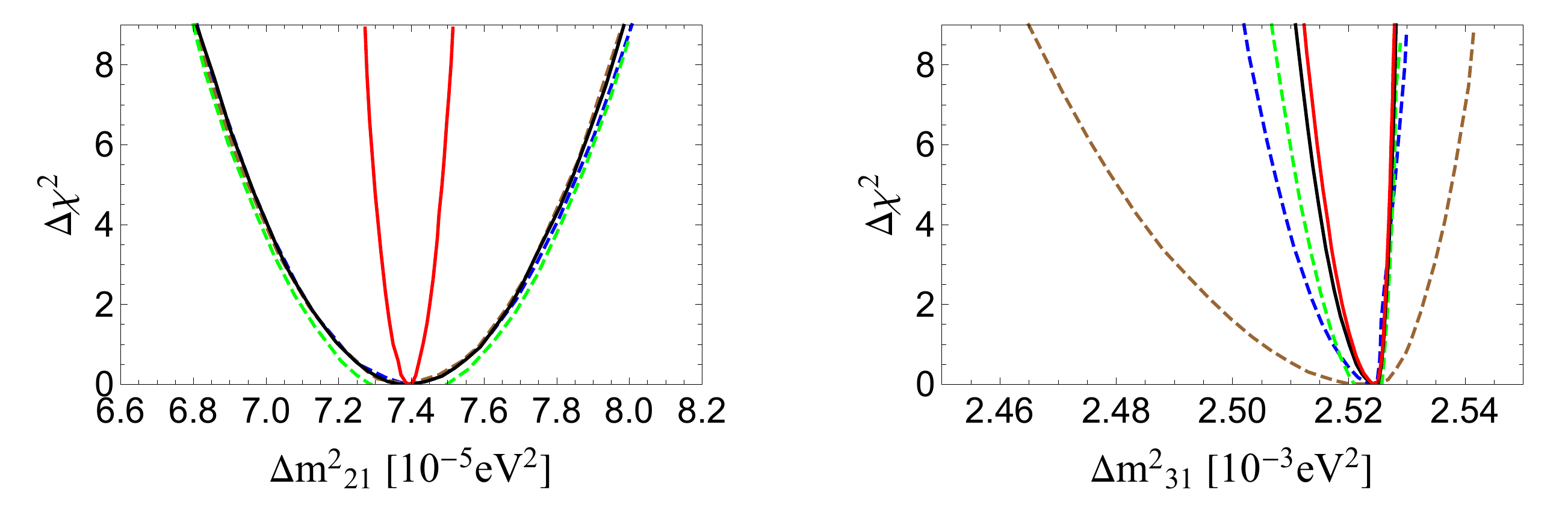}
\caption{\label{fig:inOSC_1D}The $\Delta\chi^2$ value against $\theta_{12}$ (upper left), $\theta_{13}$ (upper right), $\theta_{23}$ (middle left), $\delta$ (middle right), $\Delta m_{21}^2$ (lower left) and $\Delta m_{31}^2$ (lower right), for DUNE (dashed-blue), T2HK (dashed-green), the combination of NO$\nu$A and T2K (dashed-brown), the synergy of these four LBLs (black), and including all LBLs and JUNO (red), assuming the {tri-direct CP model}.}
\end{figure}

\begin{figure}[!ht]
 \includegraphics[width=0.45\textwidth,height=0.45\textwidth]{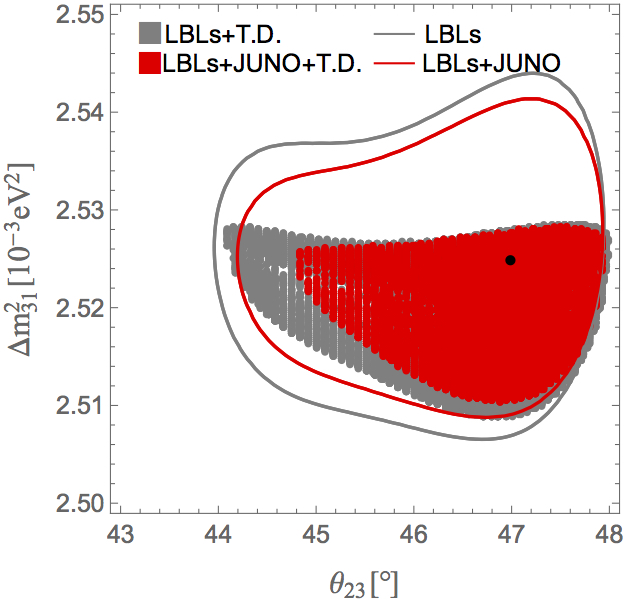}
 \includegraphics[width=0.45\textwidth,height=0.45\textwidth]{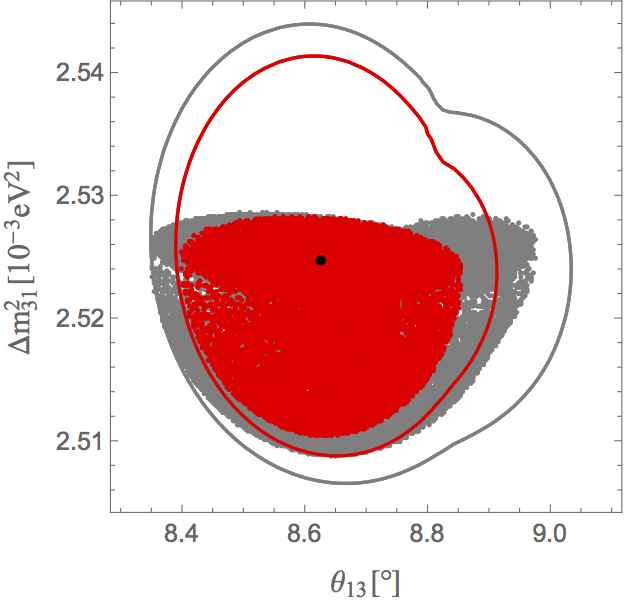}\\ \vspace{5mm}
 \includegraphics[width=0.45\textwidth,height=0.45\textwidth]{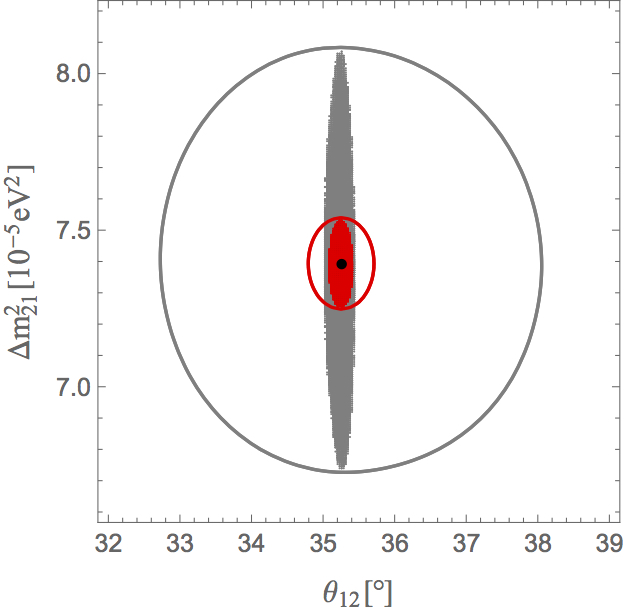}
 \includegraphics[width=0.45\textwidth,height=0.45\textwidth]{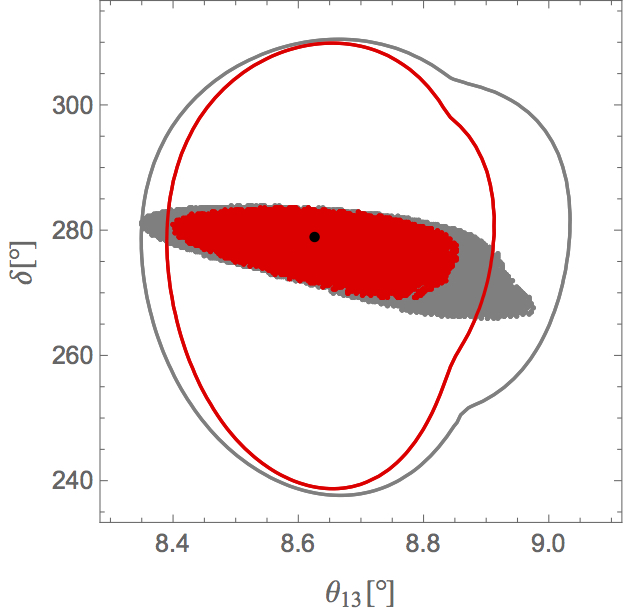}
\caption{\label{fig:inOSC_2D}The points on{ the $3\sigma$ sphere in the 4-dimension model-parameter space} projected on $\theta_{23}$-$\Delta m_{31}^2$ (upper-left), $\theta_{13}$-$\Delta m_{31}^2$(upper-right), $\theta_{12}$-$\Delta m^2_{21}$ (lower-left), $\theta_{13}$-$\delta$(lower-right) for the synergy of these four LBLs (grey), and including all LBLs and JUNO (red). We also compare these results to those without the restrictions from tri-direct models for LBLs synergy (grey contour) and combing all experiments (red contour).}
\end{figure}

In Fig.~\ref{fig:inOSC_1D}, we show $\Delta\chi^2$ values against all oscillation parameters for the combination of NO$\nu$A and T2K (brown), DUNE (dahsed-blue), T2HK (dashed-green), the synergy of these four LBLs (grey), and including all LBLs and JUNO (red), assuming the tri-direct model. We see that under the tri-direct assumption, the combination of NO$\nu$A and T2K {performs} the worst, while DUNE performs much better, except for $\theta_{13}$ and $\Delta m_{21}^2$. T2HK works slightly better than DUNE, and dominates the performance of the combination of all LBLs. The asymmetry for $\Delta m^2_{31}$ comes from the asymmetry behaviour of $x$ and $r$ in Fig.~\ref{fig:model1D} through Eq.~(\ref{eq:m2m3_TD}). Obviously, the twist behaviour for $\eta$ is passed to those for $\theta_{13}$, $\theta_{23}$ and $\delta$ by the tri-direct model. We note that even LBL experiments are not sensitive to $\theta_{12}$, the uncertainty can be improved by precisely measuring other oscillation parameters within the tri-direct model. We further point out a great improvement by including JUNO data which can be seen in the result for $\Delta m_{21}^2$. 

In Fig.~\ref{fig:inOSC_2D}, we show the points at the $3\sigma$ {surface} projected on $\theta_{23}$-$\Delta m_{31}^2$ (upper-left), $\theta_{13}$-$\Delta m_{31}^2$(upper-right), $\theta_{12}$-$\Delta m^2_{21}$ (lower-left), $\theta_{13}$-$\delta$(lower-right) for the synergy of these four LBLs (grey), and including all LBLs and JUNO (red). Because of the nonlinear relations between model parameters and standard parameters, the data do not spread uniformly. We also compare them with those without the restriction from the {tri-direct CP} model: the grey curve is for including all LBLs, while {the} dashed black {curve} is for a combination of LBLs and JUNO. There is a discontinuity when $\theta_{13}$ is larger than $\sim8.8^\circ$, because of the degeneracy with $\theta_{23}$. These results show that assuming {tri-direct CP} improves the key measurements for future experiments.

\subsection{A discrimination of two benchmark models}

\begin{figure}[!ht]
 \includegraphics[width=1.\textwidth]{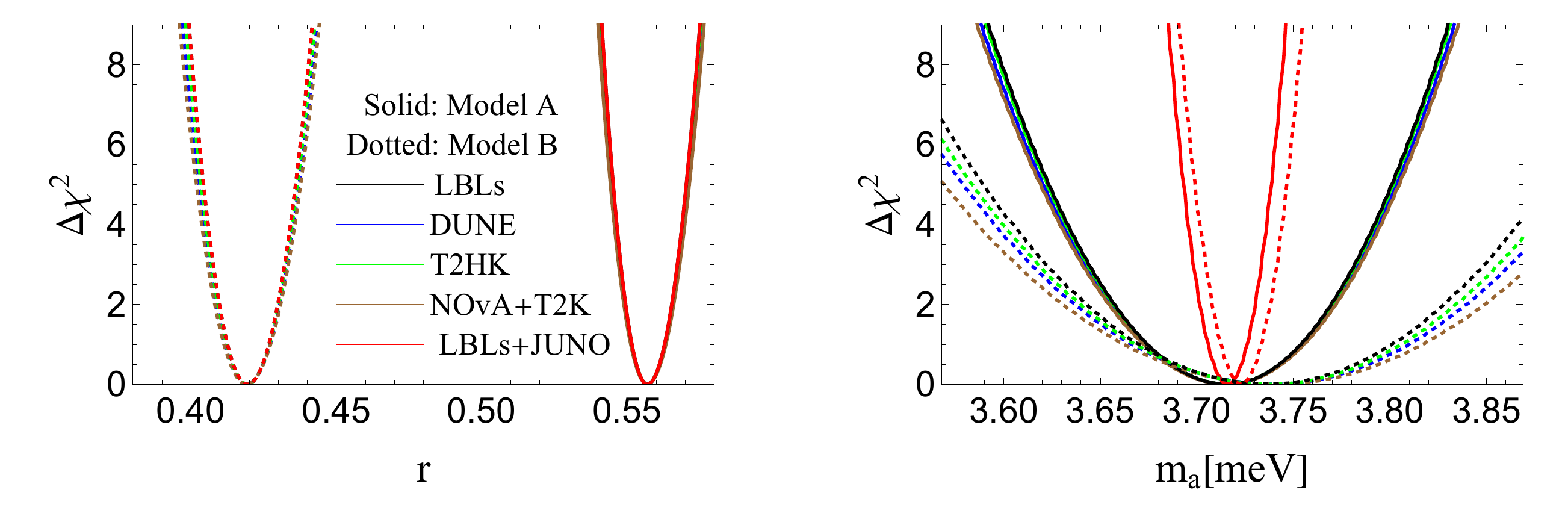}
 \caption{\label{fig:modelAB}The $\Delta\chi^2$ value against $r$ (left) and $m_a$ (right) assuming model A (solid curve; Eq.~(\ref{modelA})) and model B (dotted curve; Eq.~(\ref{modelB})), for DUNE (blue), T2HK (green), combination of NO$\nu$A and T2K (brown), the synergy of four LBLs (black) and the interplay of LBLs and JUNO (red). In the model A (B), two conditions are assumed: $x=-7/2$ and $\eta=\pi$ ($x=-4$ and $\eta=5\pi/4$), while in the current global fit results the best fit for the other two parameters are located $(r,~m_a)=(0.557,~3.716~\text{meV})$ ($(r,~m_a)=(0.421,~3.723~\text{meV})$).}
\end{figure}

In Fig.~\ref{fig:modelAB}, we show $\Delta\chi^2$ against $r$ (left) and $m_a$ (right) assuming model A (solid curve; Eq.~\eqref{modelA}) and model B (dotted curve; Eq.~\eqref{modelB}), for DUNE (blue), T2HK (green), the combination of NO$\nu$A and T2K (brown), the synergy of four LBLs (black) and the interplay of LBLs and JUNO data (red).
These two models, shown in Tab.~\ref{tab:bf_N4}, assume different values for $x$ and $\eta$: $(x,~\eta)=(-7/2,~\pi)$ for model A and $(x,~\eta)=(-4,~5\pi/4)$ for model B. {The corresponding best fits} with the global fit result are given $(r,~m_a)=(0.557,~3.716~\text{meV})$ for model A and $(r,~m_a)=(0.421,~3.723~\text{meV})$ for model B. We see that based on one model, the better way to exclude the other one is by precision measurement of $r$. The experimental configuration does not affect the uncertainty for $r$. This uncertainty is $\sim0.2$ at $3\sigma$ under both models.
Two models predict very similar values for $m_a$. As a result, it is impossible to exclude the wrong model by measuring this parameter alone. Moreover, the uncertainty of $m_a$ depends on the model and the experimental configuration. The precision under Model A is generally better than that for Model B, expect for {the combination} of LBLs and JUNO data. The rank of precision from the worst to the best experimental configuration is the combination of NO$\nu$A and T2K, DUNE, T2HK, the synergy of all LBLs, and combining all LBLs and JUNO. 
%The model parameter $r$ might not be measured directly, but there are two mass squared differences in standard neutrino mixing parameters. 
Both $m_a$ and $m_s$ will be determined precisely by experiments. As given in the definition, the model discriminator $r\equiv m_a/m_s$ points to a requirement to measure both mass squared differences in neutrino experiments as precisely as possible.

\section{Summary}
\label{sec:summary}
The tri-direct CP symmetry model offers fruitful features to accommodate neutrino masses and explain neutrino mixing and oscillations. The more powerful aspect is the model predicted correlations of standard neutrino mixing parameters preserved by an underlying symmetry. We looked into a probe of {the} tri-direct CP symmetry model by simulating the current and future neutrino oscillation experiments, including T2K, NO$\nu$A, T2HK, DUNE and JUNO. We found that the degeneracy problem cannot be avoided at a single {long baseline} experiment in the precision measurement of model parameters while a combination of long-baseline and reactor experiments will resolve the problem.
This fact highlights the complementarity of different neutrino oscillation experiments. In addition, we scanned the standard neutrino mixing parameters expressed by the underlying model ``true" values in order to {determine} how powerful precision measurements in the traditional analysis will be. It seems that shape of contours in the projected {parameter space} can {give} us hints {of the underlying theory} but the information remains limited by a multiple-channel analysis in a single experiment. {This limitation} points to a combined analysis by multiple experiments with different {beams} and baseline configurations.
Finally, we can discriminate benchmark models after a discovery of CP violation in the leptonic sector by any one of these experiments.

\section{Acknowledgement}%T.G.
This work is supported in part by the National Natural Science Foundation of China under Grant Nos. 11505301, 11881240247, 11522546 and 11835013. JT appreciates ICTP's hospitality and nice discussions with participants during the workshop PANE2018. The work was initiated and expanded at the Chinese High-Energy Physics Conference and the MOMENT\&EMuS meeting in 2018. We also thank Dr.~Nick W. Prouse to kindly provide the simulation package for T2HK. Finally, we appreciate Dr. Neil Drouard Raper's help to improve the readability of our paper.

\appendix
\section{Physics performance of different configurations}\label{app:EXP}

\begin{figure}[!ht]
 \includegraphics[width=0.15\textwidth]{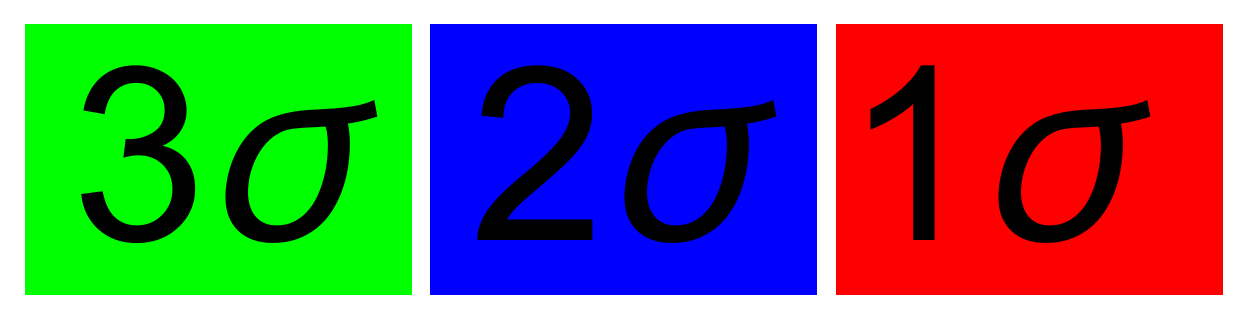}\\
 \includegraphics[width=.5\textwidth]{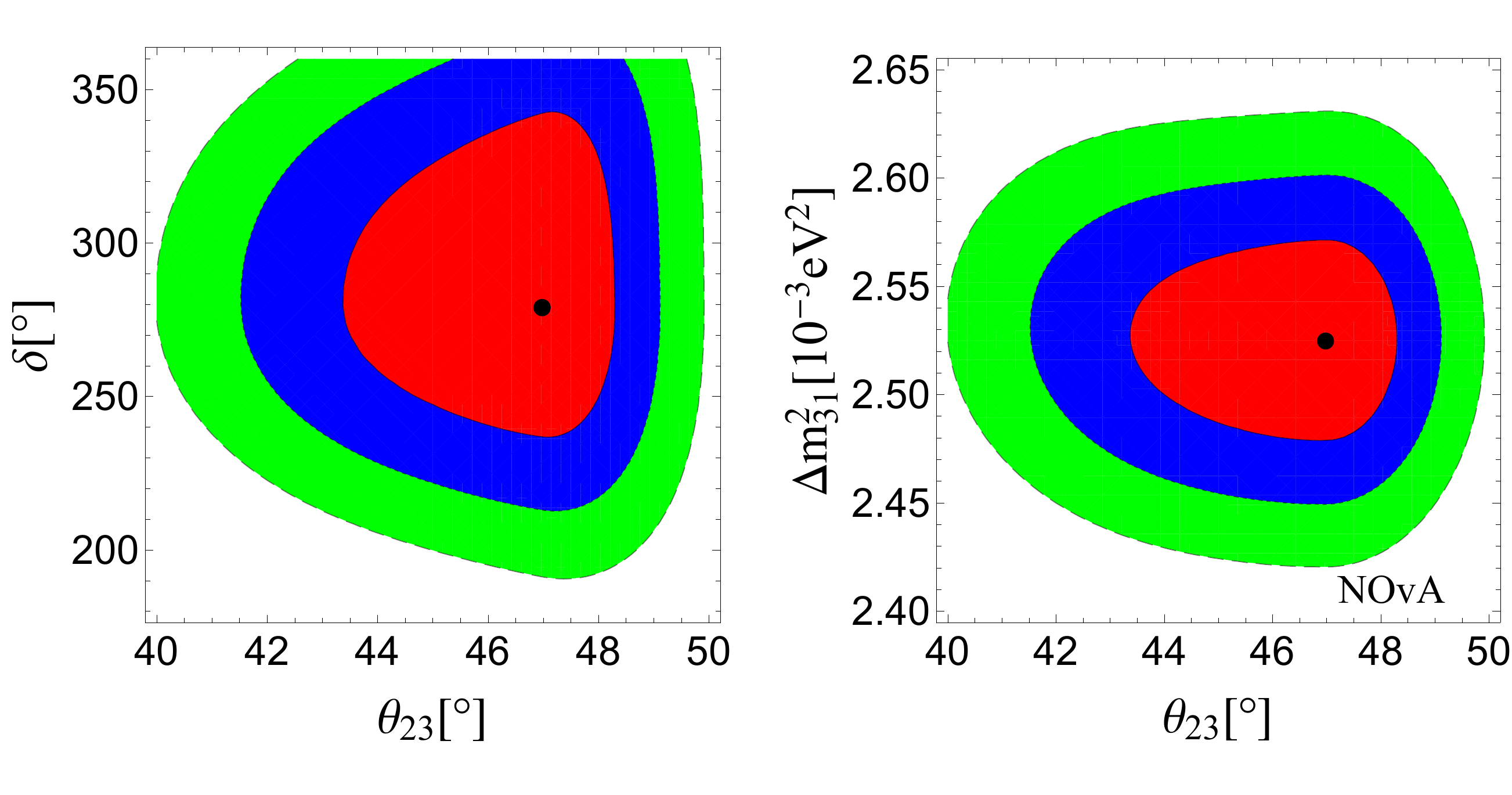}\\ \vspace{2mm}
  \includegraphics[width=.5\textwidth]{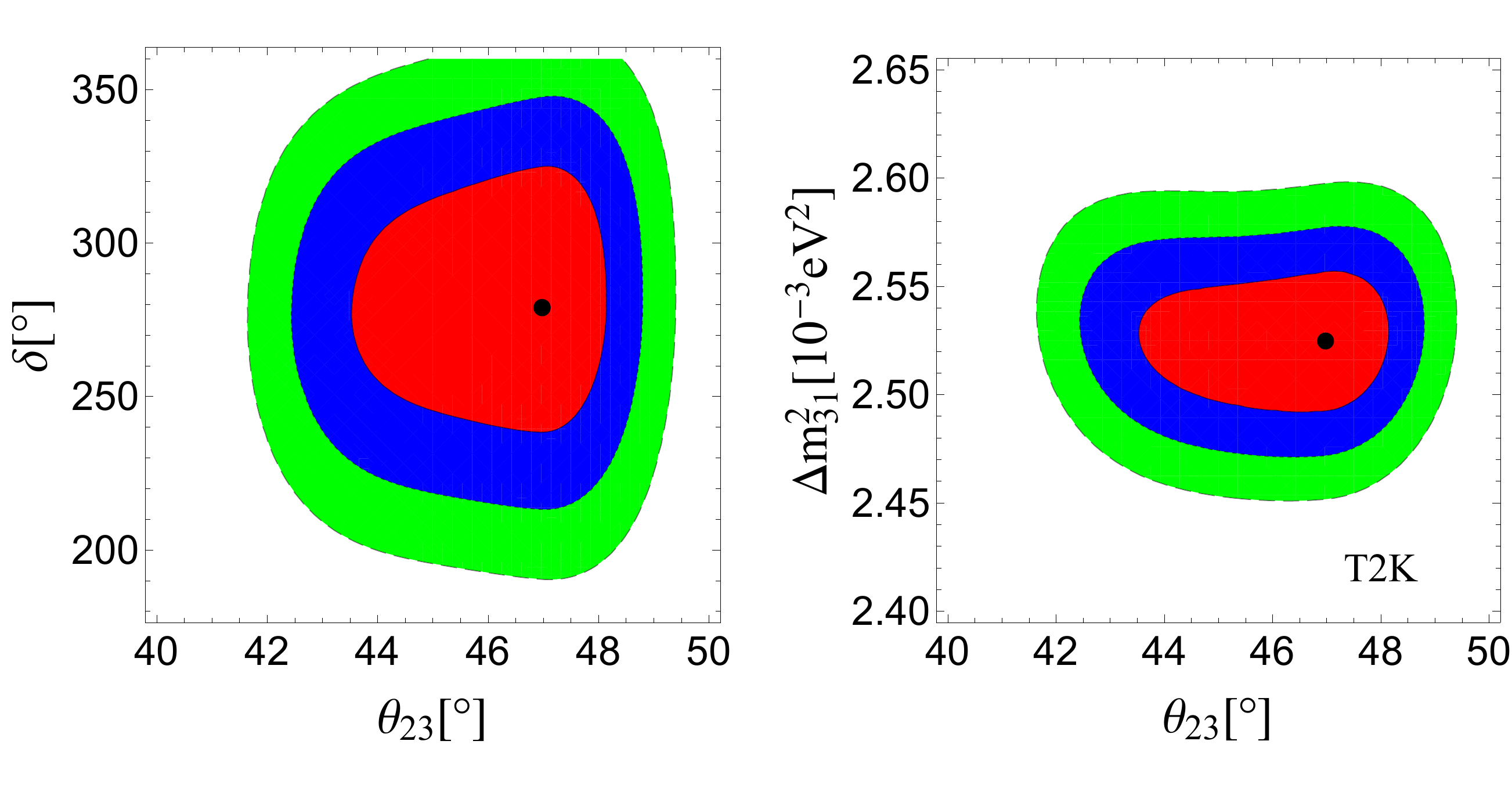}
  \includegraphics[width=.5\textwidth]{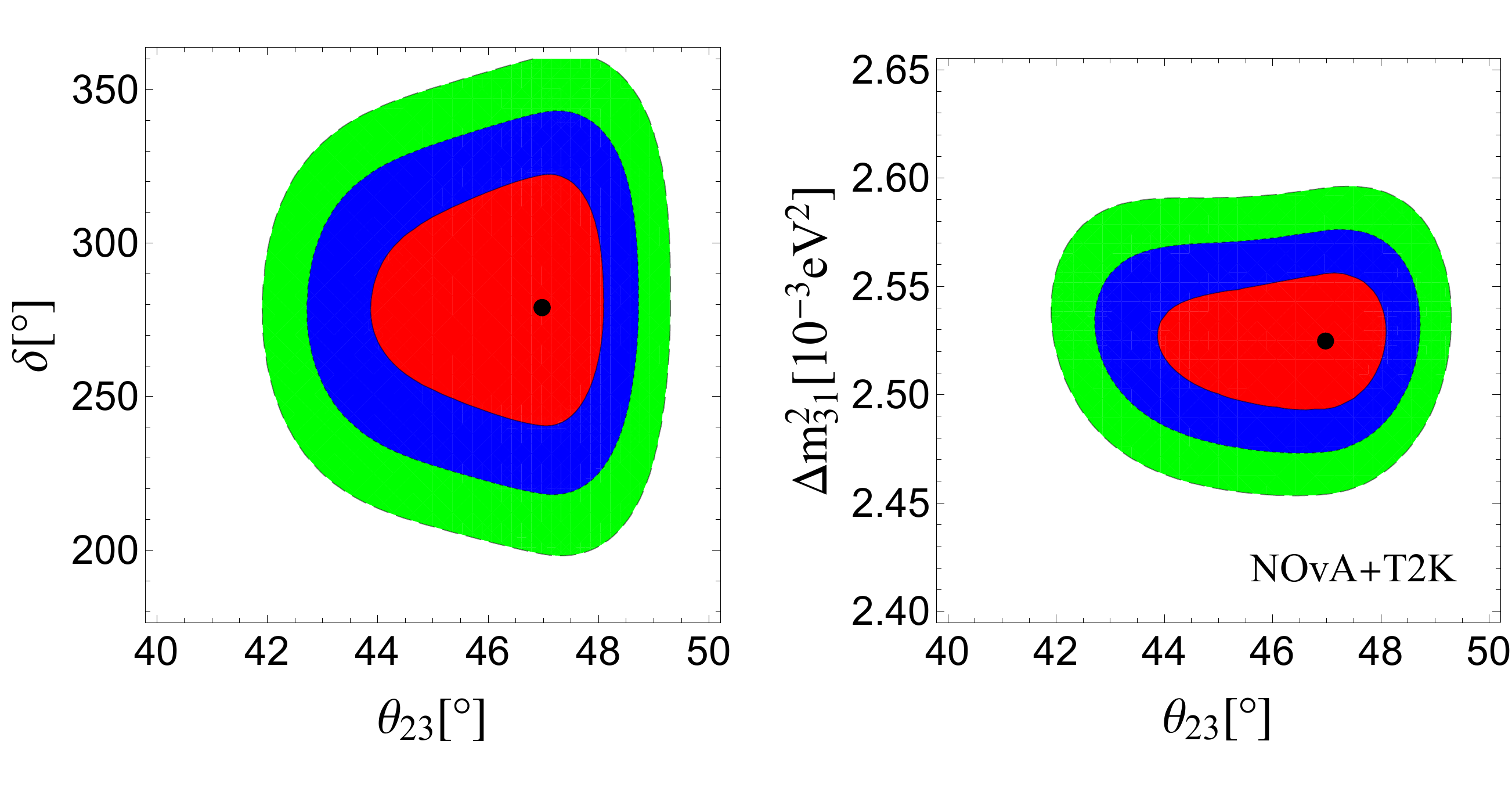}
 \caption{\label{fig:NT}The contours on $\theta_{23}$-$\delta$ (left) and $\theta_{23}$-$\Delta m_{31}^2$ (right) for NO$\nu$A (upper) planes, T2K (middle) and their combination (lower) at $1\sigma$ (red), $2\sigma$ (blue) and $3\sigma$ (green) precision. The true values are $\theta_{12}\sim35.3^\circ$, $\theta_{13}\sim8.6^\circ$, $\theta_{23}\sim47^\circ$, $\delta\sim279^\circ$, $\Delta m_{21}^2\sim 7.4\times 10^{-5}$ eV$^2$, and $\Delta m_{31}^2\sim 2.52\times 10^{-3}$ eV$^2$. These results include the NuFit4.0 results as priors.}
\end{figure}

\begin{figure}[!ht]
 \includegraphics[width=0.15\textwidth]{legend3.pdf}\\
 \includegraphics[width=0.5\textwidth]{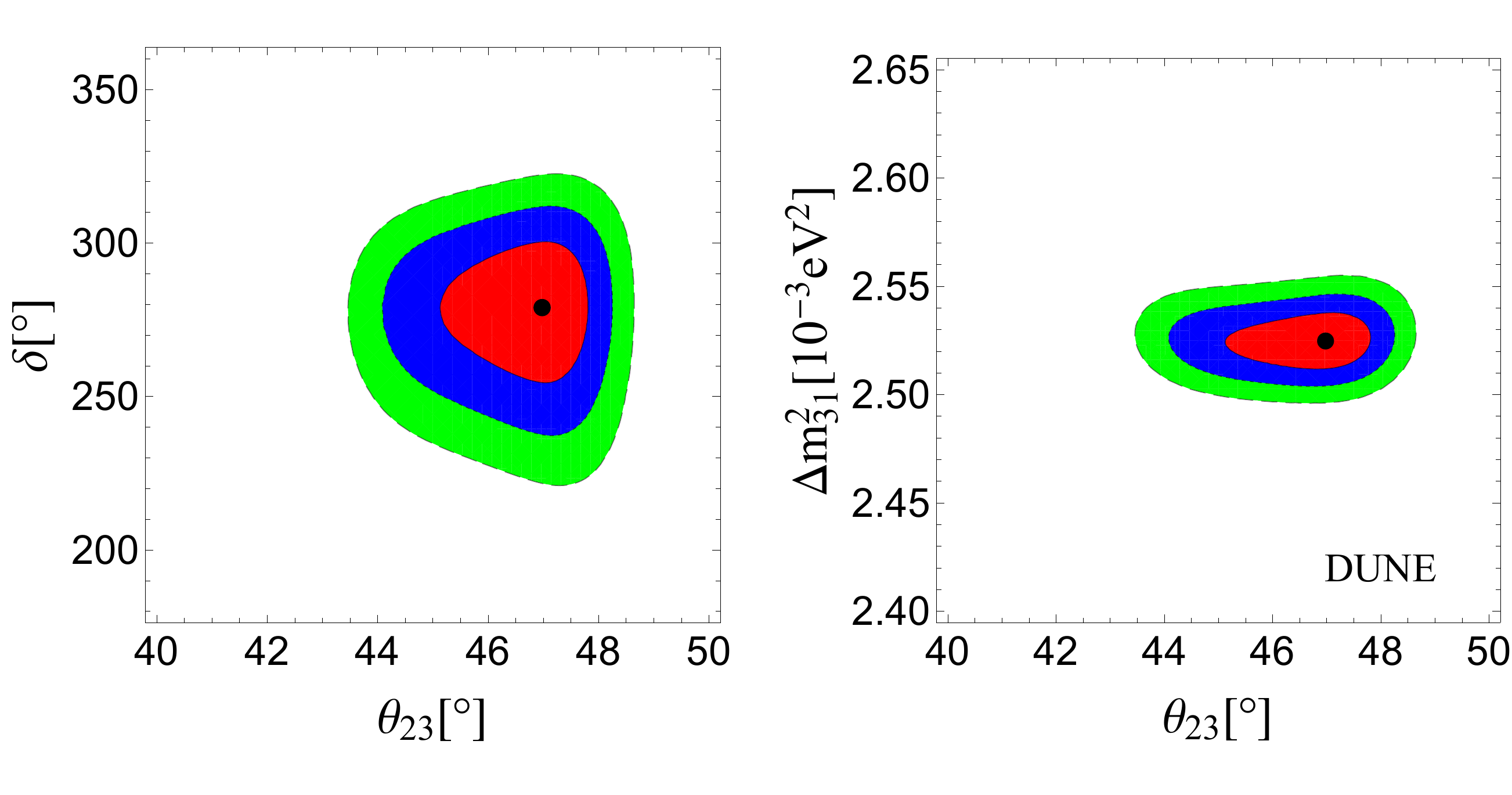}\\
 \includegraphics[width=0.5\textwidth]{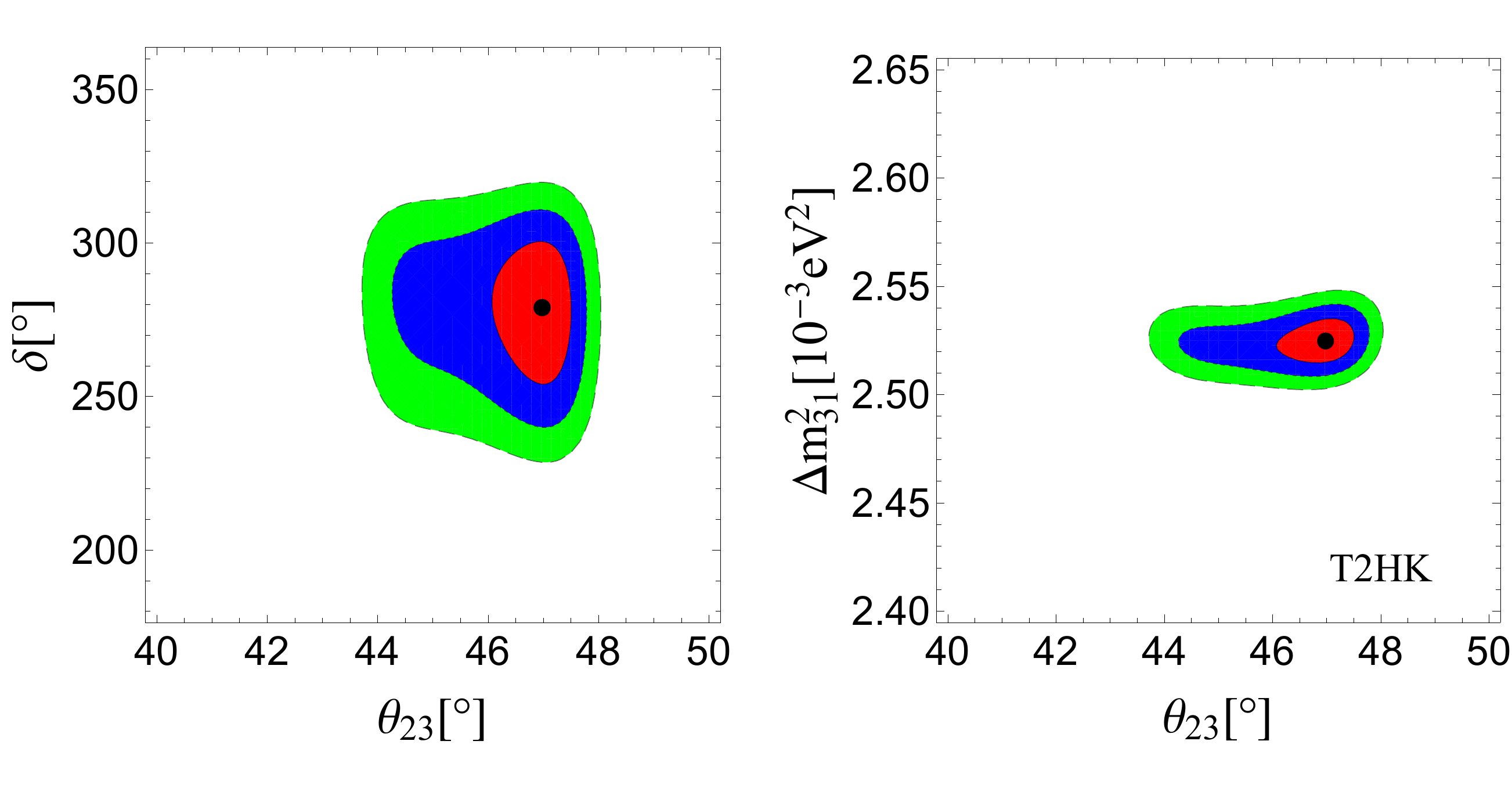}
 \caption{\label{fig:DUNE_T2HK}The contours on $\theta_{23}$-$\delta$ (left) and $\theta_{23}$-$\Delta m_{31}^2$ (right) for DUNE (upper) and T2HK (lower) at $1\sigma$ (red), $2\sigma$ (blue) and $3\sigma$ (green) precision. The true values are $\theta_{12}\sim35.3^\circ$, $\theta_{13}\sim8.6^\circ$, $\theta_{23}\sim47^\circ$, $\delta\sim279^\circ$, $\Delta m_{21}^2\sim 7.4\times 10^{-5}$ eV$^2$, and $\Delta m_{31}^2\sim 2.52\times 10^{-3}$ eV$^2$. These results include the NuFit4.0 results as priors.}
\end{figure}

In this section, we {demonstrate} the experimental potential for different configurations by showing the $1\sigma$, $2\sigma$ and $3\sigma$ {contours} on $\theta_{23}$-$\delta$ and $\theta_{23}$-$\Delta m_{31}^2$ planes. These measurements are the main goals for current and future LBLs. For current running experiments NO$\nu$A (upper) and T2K (middle), we show their expected final performance in Fig.~\ref{fig:NT}. For NO$\nu$A, we assume total $36\times 10^{20}$ POT for $\nu$ and $\bar{\nu}$ modes until $2024$, while for T2K we equally split $7.8\times10^{21}$ POT into two modes. We also show the combination of these two experiments in the lower panels of Fig.~\ref{fig:NT}. In Fig.~\ref{fig:DUNE_T2HK}, we show the performance of DUNE (upper) and T2HK (lower). For DUNE, we consider the 3-horn-optimised design with $1.83\times10^{21}$ POT per year, and we adopt $3.5$ years for each mode. For T2HK, we assume a $1.3$ MW proton beam for the neutrino source, and run $\nu$ and $\bar{\nu}$ modes for $2.5$ and $7.5$ years respectively. More details about these experiments can be seen in Secs.~\ref{sec:exp}.

\clearpage

% \normalem
\bibliographystyle{utphys}
\bibliography{flavor.bib}

\providecommand{\href}[2]{#2}\begingroup\raggedright\begin{thebibliography}{10}

\bibitem{Esteban:2018azc}
I.~Esteban, M.~C. Gonzalez-Garcia, A.~Hernandez-Cabezudo, M.~Maltoni, and
  T.~Schwetz, ``{Global analysis of three-flavour neutrino oscillations:
  synergies and tensions in the determination of $\theta_23, \delta_CP$, and
  the mass ordering},'' \href{http://dx.doi.org/10.1007/JHEP01(2019)106}{{\em
  JHEP} {\bfseries 01} (2019) 106},
\href{http://arxiv.org/abs/1811.05487}{{\ttfamily arXiv:1811.05487 [hep-ph]}}.
%%CITATION = ARXIV:1811.05487;%%.

\bibitem{Adey:2018zwh}
{\bfseries Daya Bay} Collaboration, D.~Adey {\em et~al.}, ``{Measurement of the
  Electron Antineutrino Oscillation with 1958 Days of Operation at Daya Bay},''
  \href{http://dx.doi.org/10.1103/PhysRevLett.121.241805}{{\em Phys. Rev.
  Lett.} {\bfseries 121} no.~24, (2018) 241805},
\href{http://arxiv.org/abs/1809.02261}{{\ttfamily arXiv:1809.02261 [hep-ex]}}.
%%CITATION = ARXIV:1809.02261;%%.

\bibitem{DoubleChooz:2019qbj}
{\bfseries Double Chooz} Collaboration, H.~De~Kerret {\em et~al.}, ``{First
  Double Chooz $\mathbf{\theta_{13}}$ Measurement via Total Neutron Capture
  Detection},''
\href{http://arxiv.org/abs/1901.09445}{{\ttfamily arXiv:1901.09445 [hep-ex]}}.
%%CITATION = ARXIV:1901.09445;%%.

\bibitem{Abe:2018wpn}
{\bfseries T2K} Collaboration, K.~Abe {\em et~al.}, ``{Search for CP violation
  in Neutrino and Antineutrino Oscillations by the T2K experiment with
  $2.2\times10^{21}$ protons on target},''
\href{http://arxiv.org/abs/1807.07891}{{\ttfamily arXiv:1807.07891 [hep-ex]}}.
%%CITATION = ARXIV:1807.07891;%%.

\bibitem{Cleveland:1998nv}
B.~T. Cleveland, T.~Daily, R.~Davis, Jr., J.~R. Distel, K.~Lande, C.~K. Lee,
  P.~S. Wildenhain, and J.~Ullman, ``{Measurement of the solar electron
  neutrino flux with the Homestake chlorine detector},''
\href{http://dx.doi.org/10.1086/305343}{{\em Astrophys. J.} {\bfseries 496}
  (1998) 505--526}.
%%CITATION = ASJOA,496,505;%%.

\bibitem{Abe:2010hy}
{\bfseries Super-Kamiokande} Collaboration, K.~Abe {\em et~al.}, ``{Solar
  neutrino results in Super-Kamiokande-III},''
  \href{http://dx.doi.org/10.1103/PhysRevD.83.052010}{{\em Phys. Rev.}
  {\bfseries D83} (2011) 052010},
\href{http://arxiv.org/abs/1010.0118}{{\ttfamily arXiv:1010.0118 [hep-ex]}}.
%%CITATION = ARXIV:1010.0118;%%.

\bibitem{Aharmim:2011vm}
{\bfseries SNO} Collaboration, B.~Aharmim {\em et~al.}, ``{Combined Analysis of
  all Three Phases of Solar Neutrino Data from the Sudbury Neutrino
  Observatory},'' \href{http://dx.doi.org/10.1103/PhysRevC.88.025501}{{\em
  Phys. Rev.} {\bfseries C88} (2013) 025501},
\href{http://arxiv.org/abs/1109.0763}{{\ttfamily arXiv:1109.0763 [nucl-ex]}}.
%%CITATION = ARXIV:1109.0763;%%.

\bibitem{Ghiano:2019cjy}
{\bfseries Borexino} Collaboration, C.~Ghiano and Ghiano, ``{Solar Neutrino
  Results and Future Opportunities with Borexino},''
\href{http://dx.doi.org/10.1088/1742-6596/1137/1/012054}{{\em J. Phys. Conf.
  Ser.} {\bfseries 1137} no.~1, (2019) 012054}.
%%CITATION = 00462,1137,012054;%%.

\bibitem{NOvA:2018gge}
{\bfseries NOvA} Collaboration, M.~A. Acero {\em et~al.}, ``{New constraints on
  oscillation parameters from $\nu_e$ appearance and $\nu_\mu$ disappearance in
  the NOvA experiment},'' {\em Submitted to: Phys. Rev. D} (2018) ,
\href{http://arxiv.org/abs/1806.00096}{{\ttfamily arXiv:1806.00096 [hep-ex]}}.
%%CITATION = ARXIV:1806.00096;%%.

\bibitem{Adamson:2013ue}
{\bfseries MINOS} Collaboration, P.~Adamson {\em et~al.}, ``{Electron neutrino
  and antineutrino appearance in the full MINOS data sample},''
  \href{http://dx.doi.org/10.1103/PhysRevLett.110.171801}{{\em Phys. Rev.
  Lett.} {\bfseries 110} no.~17, (2013) 171801},
\href{http://arxiv.org/abs/1301.4581}{{\ttfamily arXiv:1301.4581 [hep-ex]}}.
%%CITATION = ARXIV:1301.4581;%%.

\bibitem{Altarelli:2010gt}
G.~Altarelli and F.~Feruglio, ``{Discrete Flavor Symmetries and Models of
  Neutrino Mixing},'' \href{http://dx.doi.org/10.1103/RevModPhys.82.2701}{{\em
  Rev. Mod. Phys.} {\bfseries 82} (2010) 2701--2729},
\href{http://arxiv.org/abs/1002.0211}{{\ttfamily arXiv:1002.0211 [hep-ph]}}.
%%CITATION = ARXIV:1002.0211;%%.

\bibitem{Ishimori:2010au}
H.~Ishimori, T.~Kobayashi, H.~Ohki, Y.~Shimizu, H.~Okada, and M.~Tanimoto,
  ``{Non-Abelian Discrete Symmetries in Particle Physics},''
  \href{http://dx.doi.org/10.1143/PTPS.183.1}{{\em Prog. Theor. Phys. Suppl.}
  {\bfseries 183} (2010) 1--163},
\href{http://arxiv.org/abs/1003.3552}{{\ttfamily arXiv:1003.3552 [hep-th]}}.
%%CITATION = ARXIV:1003.3552;%%.

\bibitem{King:2013eh}
S.~F. King and C.~Luhn, ``{Neutrino Mass and Mixing with Discrete Symmetry},''
  \href{http://dx.doi.org/10.1088/0034-4885/76/5/056201}{{\em Rept. Prog.
  Phys.} {\bfseries 76} (2013) 056201},
\href{http://arxiv.org/abs/1301.1340}{{\ttfamily arXiv:1301.1340 [hep-ph]}}.
%%CITATION = ARXIV:1301.1340;%%.

\bibitem{King:2014nza}
S.~F. King, A.~Merle, S.~Morisi, Y.~Shimizu, and M.~Tanimoto, ``{Neutrino Mass
  and Mixing: from Theory to Experiment},''
  \href{http://dx.doi.org/10.1088/1367-2630/16/4/045018}{{\em New J. Phys.}
  {\bfseries 16} (2014) 045018},
\href{http://arxiv.org/abs/1402.4271}{{\ttfamily arXiv:1402.4271 [hep-ph]}}.
%%CITATION = ARXIV:1402.4271;%%.

\bibitem{King:2015aea}
S.~F. King, ``{Models of Neutrino Mass, Mixing and CP Violation},''
  \href{http://dx.doi.org/10.1088/0954-3899/42/12/123001}{{\em J. Phys.}
  {\bfseries G42} (2015) 123001},
\href{http://arxiv.org/abs/1510.02091}{{\ttfamily arXiv:1510.02091 [hep-ph]}}.
%%CITATION = ARXIV:1510.02091;%%.

\bibitem{Li:2017abz}
C.-C. Li, J.-N. Lu, and G.-J. Ding, ``{Toward a unified interpretation of quark
  and lepton mixing from flavor and CP symmetries},''
  \href{http://dx.doi.org/10.1007/JHEP02(2018)038}{{\em JHEP} {\bfseries 02}
  (2018) 038},
\href{http://arxiv.org/abs/1706.04576}{{\ttfamily arXiv:1706.04576 [hep-ph]}}.
%%CITATION = ARXIV:1706.04576;%%.

\bibitem{Lu:2018oxc}
J.-N. Lu and G.-J. Ding, ``{Quark and lepton mixing patterns from a common
  discrete flavor symmetry with a generalized CP symmetry},''
  \href{http://dx.doi.org/10.1103/PhysRevD.98.055011}{{\em Phys. Rev.}
  {\bfseries D98} no.~5, (2018) 055011},
\href{http://arxiv.org/abs/1806.02301}{{\ttfamily arXiv:1806.02301 [hep-ph]}}.
%%CITATION = ARXIV:1806.02301;%%.

\bibitem{Lu:2019gqp}
J.-N. Lu and G.-J. Ding, ``{Dihedral flavor group as the key to understand
  quark and lepton flavor mixing},''
  \href{http://dx.doi.org/10.1007/JHEP03(2019)056}{{\em JHEP} {\bfseries 03}
  (2019) 056},
\href{http://arxiv.org/abs/1901.07414}{{\ttfamily arXiv:1901.07414 [hep-ph]}}.
%%CITATION = ARXIV:1901.07414;%%.

\bibitem{Ding:2018fyz}
G.-J. Ding, S.~F. King, and C.-C. Li, ``{Tri-Direct CP in the Littlest Seesaw
  Playground},'' \href{http://dx.doi.org/10.1007/JHEP12(2018)003}{{\em JHEP}
  {\bfseries 12} (2018) 003},
\href{http://arxiv.org/abs/1807.07538}{{\ttfamily arXiv:1807.07538 [hep-ph]}}.
%%CITATION = ARXIV:1807.07538;%%.

\bibitem{Ding:2018tuj}
G.-J. Ding, S.~F. King, and C.-C. Li, ``{Lepton mixing predictions from $S_4$
  in the tridirect CP approach to two right-handed neutrino models},''
  \href{http://dx.doi.org/10.1103/PhysRevD.99.075035}{{\em Phys. Rev.}
  {\bfseries D99} no.~7, (2019) 075035},
\href{http://arxiv.org/abs/1811.12340}{{\ttfamily arXiv:1811.12340 [hep-ph]}}.
%%CITATION = ARXIV:1811.12340;%%.

\bibitem{Tanabashi:2018oca}
{\bfseries Particle Data Group} Collaboration, M.~Tanabashi {\em et~al.},
  ``{Review of Particle Physics},''
\href{http://dx.doi.org/10.1103/PhysRevD.98.030001}{{\em Phys. Rev.} {\bfseries
  D98} no.~3, (2018) 030001}.
%%CITATION = PHRVA,D98,030001;%%.

\bibitem{Wang:2018dwk}
T.~Wang and Y.-L. Zhou, ``{Neutrino nonstandard interactions as a portal to
  test flavor symmetries},''
  \href{http://dx.doi.org/10.1103/PhysRevD.99.035039}{{\em Phys. Rev.}
  {\bfseries D99} no.~3, (2019) 035039},
\href{http://arxiv.org/abs/1801.05656}{{\ttfamily arXiv:1801.05656 [hep-ph]}}.
%%CITATION = ARXIV:1801.05656;%%.

\bibitem{Esteban:2018ppq}
I.~Esteban, M.~C. Gonzalez-Garcia, M.~Maltoni, I.~Martinez-Soler, and
  J.~Salvado, ``{Updated Constraints on Non-Standard Interactions from Global
  Analysis of Oscillation Data},''
  \href{http://dx.doi.org/10.1007/JHEP08(2018)180}{{\em JHEP} {\bfseries 08}
  (2018) 180},
\href{http://arxiv.org/abs/1805.04530}{{\ttfamily arXiv:1805.04530 [hep-ph]}}.
%%CITATION = ARXIV:1805.04530;%%.

\bibitem{Tang:2017khg}
J.~Tang, Y.~Zhang, and Y.-F. Li, ``{Probing Direct and Indirect Unitarity
  Violation in Future Accelerator Neutrino Facilities},''
  \href{http://dx.doi.org/10.1016/j.physletb.2017.09.055}{{\em Phys. Lett.}
  {\bfseries B774} (2017) 217--224},
\href{http://arxiv.org/abs/1708.04909}{{\ttfamily arXiv:1708.04909 [hep-ph]}}.
%%CITATION = ARXIV:1708.04909;%%.

\bibitem{Tang:2017qen}
J.~Tang and Y.~Zhang, ``{Study of nonstandard charged-current interactions at
  the MOMENT experiment},''
  \href{http://dx.doi.org/10.1103/PhysRevD.97.035018}{{\em Phys. Rev.}
  {\bfseries D97} no.~3, (2018) 035018},
\href{http://arxiv.org/abs/1705.09500}{{\ttfamily arXiv:1705.09500 [hep-ph]}}.
%%CITATION = ARXIV:1705.09500;%%.

\bibitem{Ascencio-Sosa:2018lbk}
M.~V. Ascencio-Sosa, A.~M. Calatayud-Cadenillas, A.~M. Gago, and
  J.~Jones-Pérez, ``{Matter effects in neutrino visible decay at future
  long-baseline experiments},''
  \href{http://dx.doi.org/10.1140/epjc/s10052-018-6276-0}{{\em Eur. Phys. J.}
  {\bfseries C78} no.~10, (2018) 809},
\href{http://arxiv.org/abs/1805.03279}{{\ttfamily arXiv:1805.03279 [hep-ph]}}.
%%CITATION = ARXIV:1805.03279;%%.

\bibitem{Tang:2018rer}
J.~Tang, T.-C. Wang, and Y.~Zhang, ``{Invisible neutrino decays at the MOMENT
  experiment},''
\href{http://arxiv.org/abs/1811.05623}{{\ttfamily arXiv:1811.05623 [hep-ph]}}.
%%CITATION = ARXIV:1811.05623;%%.

\bibitem{Coloma:2015kiu}
P.~Coloma, ``{Non-Standard Interactions in propagation at the Deep Underground
  Neutrino Experiment},'' \href{http://dx.doi.org/10.1007/JHEP03(2016)016}{{\em
  JHEP} {\bfseries 03} (2016) 016},
\href{http://arxiv.org/abs/1511.06357}{{\ttfamily arXiv:1511.06357 [hep-ph]}}.
%%CITATION = ARXIV:1511.06357;%%.

\bibitem{Liao:2016orc}
J.~Liao, D.~Marfatia, and K.~Whisnant, ``{Nonstandard neutrino interactions at
  DUNE, T2HK and T2HKK},''
  \href{http://dx.doi.org/10.1007/JHEP01(2017)071}{{\em JHEP} {\bfseries 01}
  (2017) 071},
\href{http://arxiv.org/abs/1612.01443}{{\ttfamily arXiv:1612.01443 [hep-ph]}}.
%%CITATION = ARXIV:1612.01443;%%.

\bibitem{Masud:2017bcf}
M.~Masud, M.~Bishai, and P.~Mehta, ``{Extricating New Physics Scenarios at DUNE
  with Higher Energy Beams},''
  \href{http://dx.doi.org/10.1038/s41598-018-36790-6}{{\em Sci. Rep.}
  {\bfseries 9} no.~1, (2019) 352},
\href{http://arxiv.org/abs/1704.08650}{{\ttfamily arXiv:1704.08650 [hep-ph]}}.
%%CITATION = ARXIV:1704.08650;%%.

\bibitem{Kuchibhatla:2018grr}
N.~D. Kuchibhatla, S.~Goswami, and N.~Nath, ``{Effect of non-standard neutrino
  interactions on the sensitivities of DUNE},''
\href{http://dx.doi.org/10.22323/1.295.0162}{{\em PoS} {\bfseries NuFact2017}
  (2018) 162}.
%%CITATION = POSCI,NuFact2017,162;%%.

\bibitem{Deepthi:2017gxg}
K.~N. Deepthi, S.~Goswami, and N.~Nath, ``{Challenges posed by non-standard
  neutrino interactions in the determination of $\delta_{CP}$ at DUNE},''
  \href{http://dx.doi.org/10.1016/j.nuclphysb.2018.09.004}{{\em Nucl. Phys.}
  {\bfseries B936} (2018) 91--105},
\href{http://arxiv.org/abs/1711.04840}{{\ttfamily arXiv:1711.04840 [hep-ph]}}.
%%CITATION = ARXIV:1711.04840;%%.

\bibitem{Abe:2011sj}
{\bfseries T2K} Collaboration, K.~Abe {\em et~al.}, ``{Indication of Electron
  Neutrino Appearance from an Accelerator-produced Off-axis Muon Neutrino
  Beam},'' \href{http://dx.doi.org/10.1103/PhysRevLett.107.041801}{{\em Phys.
  Rev. Lett.} {\bfseries 107} (2011) 041801},
\href{http://arxiv.org/abs/1106.2822}{{\ttfamily arXiv:1106.2822 [hep-ex]}}.
%%CITATION = ARXIV:1106.2822;%%.

\bibitem{Adamson:2016tbq}
{\bfseries NOvA} Collaboration, P.~Adamson {\em et~al.}, ``{First measurement
  of electron neutrino appearance in NOvA},''
  \href{http://dx.doi.org/10.1103/PhysRevLett.116.151806}{{\em Phys. Rev.
  Lett.} {\bfseries 116} no.~15, (2016) 151806},
\href{http://arxiv.org/abs/1601.05022}{{\ttfamily arXiv:1601.05022 [hep-ex]}}.
%%CITATION = ARXIV:1601.05022;%%.

\bibitem{Adamson:2016xxw}
{\bfseries NOvA} Collaboration, P.~Adamson {\em et~al.}, ``{First measurement
  of muon-neutrino disappearance in NOvA},''
  \href{http://dx.doi.org/10.1103/PhysRevD.93.051104}{{\em Phys. Rev.}
  {\bfseries D93} no.~5, (2016) 051104},
\href{http://arxiv.org/abs/1601.05037}{{\ttfamily arXiv:1601.05037 [hep-ex]}}.
%%CITATION = ARXIV:1601.05037;%%.

\bibitem{Alion:2016uaj}
{\bfseries DUNE} Collaboration, T.~Alion {\em et~al.}, ``{Experiment Simulation
  Configurations Used in DUNE CDR},''
\href{http://arxiv.org/abs/1606.09550}{{\ttfamily arXiv:1606.09550
  [physics.ins-det]}}.
%%CITATION = ARXIV:1606.09550;%%.

\bibitem{Abe:2018uyc}
{\bfseries Hyper-Kamiokande} Collaboration, K.~Abe {\em et~al.},
  ``{Hyper-Kamiokande Design Report},''
\href{http://arxiv.org/abs/1805.04163}{{\ttfamily arXiv:1805.04163
  [physics.ins-det]}}.
%%CITATION = ARXIV:1805.04163;%%.

\bibitem{Esteban:2016qun}
I.~Esteban, M.~C. Gonzalez-Garcia, M.~Maltoni, I.~Martinez-Soler, and
  T.~Schwetz, ``{Updated fit to three neutrino mixing: exploring the
  accelerator-reactor complementarity},''
  \href{http://dx.doi.org/10.1007/JHEP01(2017)087}{{\em JHEP} {\bfseries 01}
  (2017) 087},
\href{http://arxiv.org/abs/1611.01514}{{\ttfamily arXiv:1611.01514 [hep-ph]}}.
%%CITATION = ARXIV:1611.01514;%%.

\bibitem{Abe:2017uxa}
{\bfseries T2K} Collaboration, K.~Abe {\em et~al.}, ``{Combined Analysis of
  Neutrino and Antineutrino Oscillations at T2K},''
  \href{http://dx.doi.org/10.1103/PhysRevLett.118.151801}{{\em Phys. Rev.
  Lett.} {\bfseries 118} no.~15, (2017) 151801},
\href{http://arxiv.org/abs/1701.00432}{{\ttfamily arXiv:1701.00432 [hep-ex]}}.
%%CITATION = ARXIV:1701.00432;%%.

\bibitem{Adamson:2017gxd}
{\bfseries NOvA} Collaboration, P.~Adamson {\em et~al.}, ``{Constraints on
  Oscillation Parameters from $\nu_e$ Appearance and $\nu_\mu$ Disappearance in
  NOvA},'' \href{http://dx.doi.org/10.1103/PhysRevLett.118.231801}{{\em Phys.
  Rev. Lett.} {\bfseries 118} no.~23, (2017) 231801},
\href{http://arxiv.org/abs/1703.03328}{{\ttfamily arXiv:1703.03328 [hep-ex]}}.
%%CITATION = ARXIV:1703.03328;%%.

\bibitem{Huber:2004ka}
P.~Huber, M.~Lindner, and W.~Winter, ``{Simulation of long-baseline neutrino
  oscillation experiments with GLoBES (General Long Baseline Experiment
  Simulator)},'' \href{http://dx.doi.org/10.1016/j.cpc.2005.01.003}{{\em
  Comput. Phys. Commun.} {\bfseries 167} (2005) 195},
\href{http://arxiv.org/abs/hep-ph/0407333}{{\ttfamily arXiv:hep-ph/0407333
  [hep-ph]}}.
%%CITATION = HEP-PH/0407333;%%.

\bibitem{Huber:2007ji}
P.~Huber, J.~Kopp, M.~Lindner, M.~Rolinec, and W.~Winter, ``{New features in
  the simulation of neutrino oscillation experiments with GLoBES 3.0: General
  Long Baseline Experiment Simulator},''
  \href{http://dx.doi.org/10.1016/j.cpc.2007.05.004}{{\em Comput. Phys.
  Commun.} {\bfseries 177} (2007) 432--438},
\href{http://arxiv.org/abs/hep-ph/0701187}{{\ttfamily arXiv:hep-ph/0701187
  [hep-ph]}}.
%%CITATION = HEP-PH/0701187;%%.

\bibitem{prem:1981}
A.~M. Dziewonski and D.~L. Anderson {\em Physics of the Earth and Planetary
  Interiors} {\bfseries 25} (1981) 297--356.

\bibitem{Abe:2015ibe}
{\bfseries T2K} Collaboration, K.~Abe {\em et~al.}, ``{Measurement of Muon
  Antineutrino Oscillations with an Accelerator-Produced Off-Axis Beam},''
  \href{http://dx.doi.org/10.1103/PhysRevLett.116.181801}{{\em Phys. Rev.
  Lett.} {\bfseries 116} no.~18, (2016) 181801},
\href{http://arxiv.org/abs/1512.02495}{{\ttfamily arXiv:1512.02495 [hep-ex]}}.
%%CITATION = ARXIV:1512.02495;%%.

\bibitem{Adamson:2017qqn}
{\bfseries NOvA} Collaboration, P.~Adamson {\em et~al.}, ``{Measurement of the
  neutrino mixing angle $\theta_{23}$ in NOvA},''
  \href{http://dx.doi.org/10.1103/PhysRevLett.118.151802}{{\em Phys. Rev.
  Lett.} {\bfseries 118} no.~15, (2017) 151802},
\href{http://arxiv.org/abs/1701.05891}{{\ttfamily arXiv:1701.05891 [hep-ex]}}.
%%CITATION = ARXIV:1701.05891;%%.

\bibitem{Papadimitriou:2017ytl}
V.~Papadimitriou {\em et~al.},
  \href{http://dx.doi.org/10.18429/JACoW-IPAC2016-TUPMR025}{``{Design of the
  LBNF Beamline},''} in {\em {Proceedings, 7th International Particle
  Accelerator Conference (IPAC 2016): Busan, Korea, May 8-13, 2016}},
  p.~TUPMR025.
\newblock 2016.
\newblock \href{http://arxiv.org/abs/1704.04471}{{\ttfamily arXiv:1704.04471
  [physics.acc-ph]}}.
\newblock
\url{http://lss.fnal.gov/archive/2016/conf/fermilab-conf-16-163-ad.pdf}.
\newblock
%%CITATION = ARXIV:1704.04471;%%.

\bibitem{Tariq:2016ysu}
{\bfseries LBNF} Collaboration, S.~Tariq {\em et~al.},
  \href{http://dx.doi.org/10.18429/JACoW-NAPAC2016-MOPOB35}{``{Design of the
  LBNF Beamline Target Station},''} in {\em {Proceedings, 2nd North American
  Particle Accelerator Conference (NAPAC2016): Chicago, Illinois, USA, October
  9-14, 2016}}, p.~MOPOB35.
\newblock 2017.
\newblock \href{http://arxiv.org/abs/1612.07293}{{\ttfamily arXiv:1612.07293
  [physics.acc-ph]}}.
\newblock
\url{http://lss.fnal.gov/archive/2016/conf/fermilab-conf-16-433-ad-apc-esh-nd.pdf}.
\newblock
%%CITATION = ARXIV:1612.07293;%%.

\bibitem{Djurcic:2015vqa}
{\bfseries JUNO} Collaboration, Z.~Djurcic {\em et~al.}, ``{JUNO Conceptual
  Design Report},''
\href{http://arxiv.org/abs/1508.07166}{{\ttfamily arXiv:1508.07166
  [physics.ins-det]}}.
%%CITATION = ARXIV:1508.07166;%%.

\bibitem{An:2015jdp}
{\bfseries JUNO} Collaboration, F.~An {\em et~al.}, ``{Neutrino Physics with
  JUNO},'' \href{http://dx.doi.org/10.1088/0954-3899/43/3/030401}{{\em J.
  Phys.} {\bfseries G43} no.~3, (2016) 030401},
\href{http://arxiv.org/abs/1507.05613}{{\ttfamily arXiv:1507.05613
  [physics.ins-det]}}.
%%CITATION = ARXIV:1507.05613;%%.

\bibitem{Li:2016txk}
Y.-F. Li, Y.~Wang, and Z.-z. Xing, ``{Terrestrial matter effects on reactor
  antineutrino oscillations at JUNO or RENO-50: how small is small?},''
  \href{http://dx.doi.org/10.1088/1674-1137/40/9/091001}{{\em Chin. Phys.}
  {\bfseries C40} no.~9, (2016) 091001},
\href{http://arxiv.org/abs/1605.00900}{{\ttfamily arXiv:1605.00900 [hep-ph]}}.
%%CITATION = ARXIV:1605.00900;%%.

\bibitem{Li:2018jgd}
Y.-F. Li, Z.-z. Xing, and J.-y. Zhu, ``{Indirect unitarity violation entangled
  with matter effects in reactor antineutrino oscillations},''
  \href{http://dx.doi.org/10.1016/j.physletb.2018.05.079}{{\em Phys. Lett.}
  {\bfseries B782} (2018) 578--588},
\href{http://arxiv.org/abs/1802.04964}{{\ttfamily arXiv:1802.04964 [hep-ph]}}.
%%CITATION = ARXIV:1802.04964;%%.

\end{thebibliography}\endgroup

\end{document}